\documentclass{ametsocv6.1}

\usepackage{amsmath,amsfonts,amssymb,bm}
\usepackage{mathptmx}
\usepackage{newtxtext}
\usepackage{newtxmath}

\usepackage{hyperref}
\hypersetup{colorlinks, citecolor=blue}

\title{Simulation of Atlantic Hurricane Tracks and Features: A Deep Learning Approach}

\authors{Rikhi Bose,\aff{a}\correspondingauthor{Rikhi Bose, rikhi.bose@gmail.com} 
Adam L. Pintar,\aff{b} 
and Emil Simiu,\aff{a} 
}

\affiliation{\aff{a}{Materials and Structural Systems Division, Engineering Laboratory, National Institute of Standards and Technology, Gaithersburg, MD, 20899}\\
\aff{b}{Statistical Engineering Division, Information Technology Laboratory, National Institute of Standards and Technology, Gaithersburg, MD, 20899}\\
}

\abstract{
The objective of this paper is to employ machine learning (ML) and deep learning (DL) techniques to obtain from input data (storm features) available in or derived from the HURDAT2 database models capable of simulating important hurricane properties such as landfall location and wind speed that are consistent with historical records. 
In pursuit of this objective, a trajectory model providing the storm center in terms of longitude and latitude, and intensity models providing the central pressure and maximum 1--$min$ wind speed at 10 $m$ elevation were created. 
The trajectory and intensity models are coupled and must be advanced together, six hours at a time, as the features that serve as inputs to the models at any given step depend on predictions at the previous time steps.
Once a synthetic storm database is generated, properties of interest, such as the frequencies of large wind speeds may be extracted from any part of the simulation domain. 
The coupling of the trajectory and intensity models obviates the need for an intensity decay inland of the coastline. 
Prediction results are compared to historical data, and the efficacy of the storm simulation models is demonstrated for three examples: New Orleans, Miami and Cape Hatteras. 
}

\begin{document}

\maketitle

%
%
%

%



\section{Introduction}
\label{sec1}

In thermodynamics terms, tropical storms are heat engines driven by the temperature gradient between the hot moist air close to the sea surface and the cold air in the lower stratospheric layer of the atmosphere \citep{emanuel1987dependence, emanuel1991theory}. 
This is why the probability of a storm formation is largest during the months of highest sea surface temperature. 
Due to the pressure difference between the core and the outer regions, colder air is driven towards the storm center, accumulating latent heat due to vaporization on the sea surface getting warmer and lighter, and eventually moving upward. 
The rising moist air forms the cloud systems around a storm. 
The large cloud systems rotate about their low-pressure core owing to the Coriolis effect. 
In the northern hemisphere the rotation is counterclockwise. 
Hurricanes are defined as tropical storms originating in the Atlantic basin, with maximum sustained wind speeds exceeding 74 $mph$ (33 $ms^{-1}$). 
Upon landfall, hurricanes can inflict utmost devastation. 
The economic loss due to hurricane Maria in 2017, in which more than 4000 human lives were lost \citep{kishore2018mortality}, was estimated at $\approx$ 92 billion in US dollars. 
The purpose of this paper is to present a Machine and Deep Learning (ML-DL) based methodology for the simulation of hurricane tracks and features. 

Early estimates of landfalling tropical storm wind speeds were based on (i) probabilistic models fitted to recorded translation velocities, radii of maximum wind speeds and central pressure deficits fitted to historical data on storms arriving within a chosen distance of the target location, (ii) Monte Carlo simulations of those features, and (iii) physical models of the wind speeds of interest as
functions of the simulated features \citep{russell1971probability, batts1980hurricane, georgiou1983design, neumann1987national}. 
The main weaknesses of this approach were the practical impossibility of determining reliably the tails of the probabilistic models being used, and the quality of the datasets and physical models available in the 1970s and 1980s. 

\citet{darling1991estimating} proposed a method to resolve these issues by introducing the relative storm intensity obtained by scaling the actual storm intensity by the potential intensity or the maximum possible storm intensity that mean seasonal climatic conditions would allow. 
This method is inadequate in regions where the potential intensity is small or zero. 
\citet{vickery2000simulation, vickery2009us} simulated large numbers of synthetic storms (corresponding to periods of e.g., $\approx 20,000-yr$) to generate wind maps at chosen mileposts in the U.S. coastline. 
Storm locations and intensities at $6-hr$ intervals were predicted using linear models; the storm trajectory model depended on storm latitude, longitude, translation speed and direction at two previous $6-hr$ instants, while the (relative) intensity model was a function of the storm intensities and the sea-surface temperature at two and three previous time instants, respectively. 
Different intensity models were used for easterly and westerly directions. 
The constant coefficients of the models were obtained in chosen grids discretizing the $2-D$ latitude-longitude domain. 
The synthetic storm descriptions included radial profiles of storms using the empirical profiles proposed by \citet{holland1980analytic}. 
Good agreement with the historical storm database was achieved for relevant storm parameters (central pressure, storm translation speed and heading, approach distance). 

Following \citet{vickery2000simulation}, \citet{emanuel2006statistical} also generated large numbers of storms using two statistical models for storm trajectory, and a deterministic model for storm intensity. 
One of the trajectory models utilized a Markov chain for each $6-hr$ displacement, which accounted for the $6-hr$ rate of change of direction of storm travel at the current location and time based on local climatological conditions. 
The second trajectory model also included the effect of the vertical wind shear, which also affects the intensity of a storm, and therefore, implicitly couples the trajectory and intensity models. 
Observed monthly statistics of climatological conditions were preserved while generating the synthetic tracks. 
Most importantly, instead of a statistical model, for the first time a simple physical model for storm intensity coupled with an ocean model was used \citep{emanuel2004tropical}. 
The intensity model intrinsically reduces a storm's intensity at/ after landfall. 

More recently, \citet{snaiki2020revisiting} applied the statistical-deterministic approach of \citet{emanuel2006statistical}. 
In their physics-based model for temporal intensity evolution, interplay between the inward advection of angular momentum and decay due to the frictional effects dictated the evolution of a storm's intensity; the two terms were obtained empirically from 33 hurricanes that occurred between 2001 and 2016. 
An additional decay model was necessary once a storm made landfall. 

ML and DL methods have recently achieved major advances in many areas, such as computer vision \cite{roy2022fast}, by leveraging large amounts of data to uncover relationships between feature and target variables. 
In the present work the functional dependence includes as inputs features available in or derived from the hurricane database maintained by the National Hurricane Center (NHC). 
We propose a combined ML-DL approach for the efficient simulation of storm trajectories and intensities over multiple $6-hr$ time intervals. 
The approach uses separate models for storm trajectory and intensity. 
The trajectory model provides the storm-eye location. 
Models for the central pressure and maximum $1-min$ wind speed at 10 $m$ elevation constitute the intensity model. 
Both the trajectory and storm intensity model must be advanced together, six hours at a time, as the features that serve as inputs include features from the previous prediction step. 
The proposed approach has several advantages: 
\begin{enumerate}
    \item Unlike in many of the previous works, the trajectory and intensity models are coupled, which is physically more realistic 
    \item Decay of a storm's intensity on land is naturally acquired by ML models during training, which obviates the need to model decay either by statistical or analytical means, such as in \cite{kaplan1995simple} for example
    \item The accuracy of the DL trajectory models used herein is comparable to the accuracy of ensemble models used by the NHC for track forecasting, in spite of requiring significantly less input data \citep{bose2022real}
    \item As demonstrated herein, the ML-DL models are capable of emulating historical storm patterns at small (city) and large (Atlantic basin) spatial scales 
    \item Once a synthetic storm database is generated, properties of interest, such as the probability that the wind speed exceeds 45 m/s, may be directly extracted at any location in the simulation domain 
\end{enumerate}

Recent success in applying ML/ DL methods to weather forecasting has motivated the current effort. 
In their review of emerging ML/ DL techniques for earth system science research, \citet{reichstein2019deep} drew equivalency between typical ML/ DL tasks, namely, classification, super resolution and fusion, and space- or time-dependent state prediction, and their corresponding equivalents in earth system sciences. 
Further, they pointed out the challenges to overcome for successful adoption of ML/ DL in that field. Examples are model interpretability, physical consistency, complexity, uncertainty, massive volumes of data and consequent massive computational requirements, and the dearth of labelled data. 
One of the earliest uses of machine learning for the purpose of several weather forecasting tasks was the Dynamical Integrated foreCast (DICast) \citep{myers2011consensus} that is used by the National Center for Atmospheric Research (NCAR) for forecasting variables, such as, temperature, wind speed, irradiance, etc \citep{haupt2018machine}. 
DICast is used for correcting outputs from several numerical weather prediction (NWP) models based on these models' past performance. 
\citet{kirkwood2021framework} used quantile regression forests to obtain error distributions of outputs from different NWP models. 
The probabilistic forecasts were then combined using quantile averaging that are then interpolated between aggregate quantiles to provide the final prediction distribution. 
\citet{jakaria2020smart} tested several ML algorithms, such as Ridge Regression, Support Vector, Multi-layer Perceptron (MLP), Extra-tree regression, and Random Forests (RFs) for forecasting the chosen weather conditions in the near future. 
\citet{hill2020forecasting} trained the RFs \citep{breiman1996bagging, breiman2001random}, a powerful ensemble learning ML method, on nine years of historical forecasts from NOAA's ensemble forecasts for probabilistic predictions of severe weather conditions. 
The RFs were shown to outperform the Storm Prediction Center's (SPC) outlooks on second and third day predictions, but slightly underperformed for the first day forecasts. 
RFs have also been employed by \citet{herman2018money} and \citet{kuhnlein2014improving} for predictions of precipitation, with some success. 
Applying the RFs for developing a guidance model for the purpose of severe weather forecasts based on convection-allowing ensembles, \citet{loken2020generating} concluded that the RFs provided skillful probabilistic results for severe weather prediction. 
\citet{yao2020application} applied the RFs for hail forecasting in the Shadong Peninsula; data from 41 meteorological stations over a period of 16 years were used to train the models and the model provided $6-hr$ forecasts. 
The model was found to be effective in providing accurate time of hail disasters at all hail stations. 

\citet{schultz2021can} reviewed the current workflow in numerical weather forecasting and the possible application of DL models to augment/ replace specific parts/ stages of it. 
In geosciences, the forecasting problem is formulated from satellite images, and therefore, mainly formulated as an image processing problem. 
For such problems, Convolutional Neural Networks (CNN) are known to provide superior performance \citep{lecun1998gradient}. 
In a first attempt to apply DL to climatology, \citet{liu2016application} successfully developed a CNN-based model to classify extreme weather events such as tropical cyclones, atmospheric rivers, and weather fronts using climate data from simulations and reanalysis. 
\citet{shi2018weather} introduced an edge-detection technique, and utilized a mask region-based CNN or R-CNN model to successfully predict four weather conditions -- sunny, foggy, rainy and snowy. 
\citet{pothineni2018kloudnet} proposed a CNN-based model trained on sky images for short-term forecasting of irradiance at the location of a photo-voltaic power plant to build an efficient and stable power system. 
\citet{wen2020deep} trained their CNN-based models on stacked sky images in time to extract spatio-temporal variation of cloud motion to improve intermittent photo-voltaic power generation. 
\citet{kleinert2021intellio3} built a model comprised of multiple CNN layers grouped together as inception blocks for the prediction of near-surface ozone concentrations at arbitrary air quality monitoring stations in Germany. 
The model inputs $6-day$ series data of chemical and meteorological variables to forecast up to 4 days. 

Forecasting the evolution of dynamical systems is comparable to natural language processing/ translation in DL. 
For such problems, Recurrent Neural Networks (RNN) were invented. 
\citet{gomez2003local} used a class of RNN architecture, called Long short-term memory (LSTM) \citep{hochreiter1997long} to successfully predict the maximum ozone concentration in the east Austrian region. 
\citet{qing2018hourly} also used the LSTM architecture and weather forecasting data to provide hourly day-ahead solar irradiance forecasts. 
\citet{shi2017deep} developed an accurate RNN model for short-term forecasting of precipitation based on Gated Recurrent Units (GRUs) \citep{cho2014properties} that embed the location-variant property of motion and transformation. 

For space-time dependent problems, the CNN and RNN models may be used together or in tandem. 
A Convolutional Long short-term memory (ConvLSTM), a mixed neural network model comprising CNN and  LSTM-RNN layers was used by \citet{kim2019deep} to extract spatio-temporal information from a large database of instantaneous atmospheric conditions recorded as a pixel-level history of storm tracks. 
A tensor-based Convolutional Neural Network (TCNN) was used in \cite{chen2020novel} to improve forecasts of hurricane intensity. 
The model performed a classification task that categorized storm intensity, and a regression task for wind speed estimation. 
The regression model benefited from the output of the classification model. 
Hurricane track forecasts using satellite images by \citet{ruttgers2019prediction} used a Generative Adversarial Network (GAN) \citep{goodfellow2014generative}. 
They used image time-series of typhoons in the Korean peninsula for model training. 
The GAN model was tested on 10 storms previously unseen by the model. 
The average prediction error for $6-hr$ forecasts for the test storms was 95.6 km. 
In a recent work, CNNs coupled with GRU-RNNs were employed by \cite{lian2020novel} to forecast hurricane tracks. 
A feature selection layer used before the NN layers augmented the model's learning capability from the underlying spatio-temporal structures inherent in trajectories of tropical cyclones. 
An average $12-hr$ forecast error of $\approx 100$ km between the predicted and true cyclone-eye locations, which is comparable or marginally less than the statistical model used in \cite{jeffries1993tropical} and a numerical model used in \cite{demaria1992nested} was realized. 
However, the $72-hr$ track forecast error was was less than half the errors reported for traditional track forecasting models. 
\citet{guen2020disentangling} proposed a two-branch deep neural network model named $PhyDNet$ to disentangle physical knowledge learned from partial differential equations (PDEs) from other information with unknown dynamics. 
This work could have far-reaching implications on weather/ climate forecasting that use time series data governed by PDEs, however, contaminated by noise from several sources that are not straight forward to quantify. 

The LSTM-RNN models used herein for time-marching a storm's trajectory developed by \citet{bose2021forecasting, bose2022real} have Many-To-Many prediction architecture. 
These models were tested for hundreds of validation and test storms and the prediction errors were extensively analyzed. 
Mean six- and twelve-hour forecast errors were $\approx 30$ km and 66 km, respectively. 
The $2/3$-rd probability circle radii, a statistical measure used for characterizing errors incurred for trajectory forecasting, were found for the models used herein to be comparable to the state-of-the-art ensemble models currently in use for short-term storm trajectory forecasting. 
For modelling a storm's intensity, RFs were used. 
The RFs use bagging or bootstrap aggregating. 
The considerable success RFs in applied weather forecasting for providing excellent probabilistic forecasts motivated their use here. 
The RFs require relatively less data compared to NNs and are shown to be relatively insensitive to outliers with moderate variance and low bias. Additionally, RFs have the added advantage of interpretability.

The paper is structured as follows. 
In Section~\ref{sec2}, the HURDAT2 database used for model development is discussed first from the statistical, feature engineering and model formulation points of view. 
The methodologies used for the calculation of $6-hr$ storm displacement probabilities and the modeling of storm genesis are described in this section; the types of models used, their architectures and implementation, training strategies, hyperparameter tuning as well as the simulation methodology are discussed. 
Section~\ref{sec3} presents the prediction results of the trained conjugate simulation model by comparing the global statistics of the simulated storm trajectories with those of the HURDAT2 database. 
A comparison of the predicted trajectories with those of historical test storms is also presented in this section. 
Section~\ref{sec4} demonstrates the efficacy of the synthetic storm simulation model in predicting the long-term estimates of storm wind speeds at some locations along the U.S. North Atlantic coastline; the analyses include comparisons, probability distributions of storm intensity and storm trajectories. 
Section~\ref{sec5} presents conclusions based on  our main results,  discusses the limitations, and the scope for future improvements. 

\section{Model development}
\label{sec2}

\subsection{Databases}

The National Hurricane Center conducts a post-storm analysis of each storm and updates a database that contains a six-hour best track for each storm analyzed. 
Features of the Atlantic basin storms were originally tabulated in the HURDAT (HURricane DATabase) database \citep{jarvinen1984tropical}. 
The updated version of the original database, named HURDAT2 (Hurricane Data $2^{nd}$ generation) \citep{landsea2013atlantic} is used herein. 
HURDAT2 lists the storm numbers and names, the time of the record (year, month, date and time), system status (e.g., storm classification by intensity) corresponding to the time instant, record identifiers (e.g., landfall, change of system status),  storm location (latitude and longitude), maximum $1-min$ wind speed at 10 $m$ elevation in $knots$, central pressure in millibars, and radius in nautical miles corresponding to 34, 50 and 64 knot wind speeds in all four quadrants. 
However, some database records are incomplete. 
For example, the central pressure is tabulated for each storm only since 1979, and the wind radii since 2004. 
Moreover, although storms have been tabulated since 1851, data tabulated before the use of satellites in 1970s was based on sparse observations, and seems to be less complete, i.e., some storms may not have been recorded. 
For these reasons, only part of the database is used for model training and testing.

A detailed description of the compilation of the database for model training is given by \citet{bose2021forecasting}, who developed the methodology, data compilation techniques, and model training and testing utilized herein for the trajectory models. 
The input features to the models are chosen or derived from the variables tabulated in the HURDAT2, namely, the latitude ($\phi$) and longitude ($\lambda$) coordinates, the maximum $1-min$ sustained wind speed at 10 $m$ elevation ($w_m$), central pressure ($p_c$), etc. 
Different models are trained on different sets of input features depending on the purpose of the model (see Tables \ref{t1} and \ref{t2}), so multiple distinct data sets were created from the original HURDAT2 by considering various sets of restrictions.  Only storms with six or more time records are retained, which reduces the number of storms from 1893 to 1825.  The database is further reduced to 1754 storms by retaining only the storms with $w_m$ recorded for all time instants.  By noticing that only 33 time records have $\phi > 70^\circ N$, 26 have $\lambda > 10^\circ E$, and 93 have, $V > 25$ $ms^{-1}$, 71 further storms may be excluded.  That collection of 1683 storms is referred to in what follows as DB-1, and it is used to train the LSTM-RNN trajectory models (described further in Subsection \ref{sec:trajectory-models}).


In comparison to the LSTM-RNN models, the RF models for trajectory and intensity use $p_c$ as an input feature, which is available at each time instant for 583 of the 1683 storms in DB-1.  This collection is referred to as DB-2. 
Since, for the purpose of estimating extreme wind speeds it is appropriate to focus on high intensity stoms, DB-2 is reduced further to 384 storms that attain $w_m \ge 25\, ms^{-1}$ at least once in their lifespan. 
This reduced database is named DB-3, and is used to train the RF intensity models.

\subsection{Input features}

\citet{bose2021forecasting} developed a set of LSTM-RNN models for forecasting storm trajectories.  Many of the input features for those models are used in this work, and so are reviewed for completeness.
The spherical coordinates ($\lambda$, $\phi$) are used unchanged from the database, and the storm translation direction and speed are needed for storm intensity modeling \citep{emanuel2006statistical, schwerdt1979meteorological}. 
It is assumed that the storm translation is linear between two time instants six hours apart. 
Translation direction ($\theta$) is obtained at $l^{th}$ time instant as 

\begin{equation}
\theta_l = \tan^{-1}\bigg( \frac{\phi_l -\phi_{l-1}}{\lambda_l -\lambda_{l-1}} \bigg)
\label{theta}
\end{equation}

\noindent The distance ($d$) between storm locations at two time instants was calculated using the Haversine formula, and the six-hour-averaged translation speed ($V$) is 

\begin{equation}
V_l = \frac{d(\phi_{l-1}, \phi_l, \lambda_{l-1}, \lambda_l)}{\Delta t \equiv 6 hrs}.
\label{vel}
\end{equation}

\noindent The maximum $1-min$ sustained wind speed ($w_m$) at 10 $m$ elevation is used unchanged from HURDAT2. 
In HURDAT2, $w_m$ is approximated to the nearest 5.14 $ms^{-1}$ (10 $kt$) between 1851 and 1885 and to the nearest 2.57 $ms^{-1}$ (5 $kt$) thereafter. 
The central pressure ($p_c$) in $mbar$ is used as given in the HURDAT2 database. 

\begin{figure}[!ht]
    \begin{center}
        \includegraphics[trim=0cm 0cm 0cm 0cm,clip,width=0.9\textwidth]{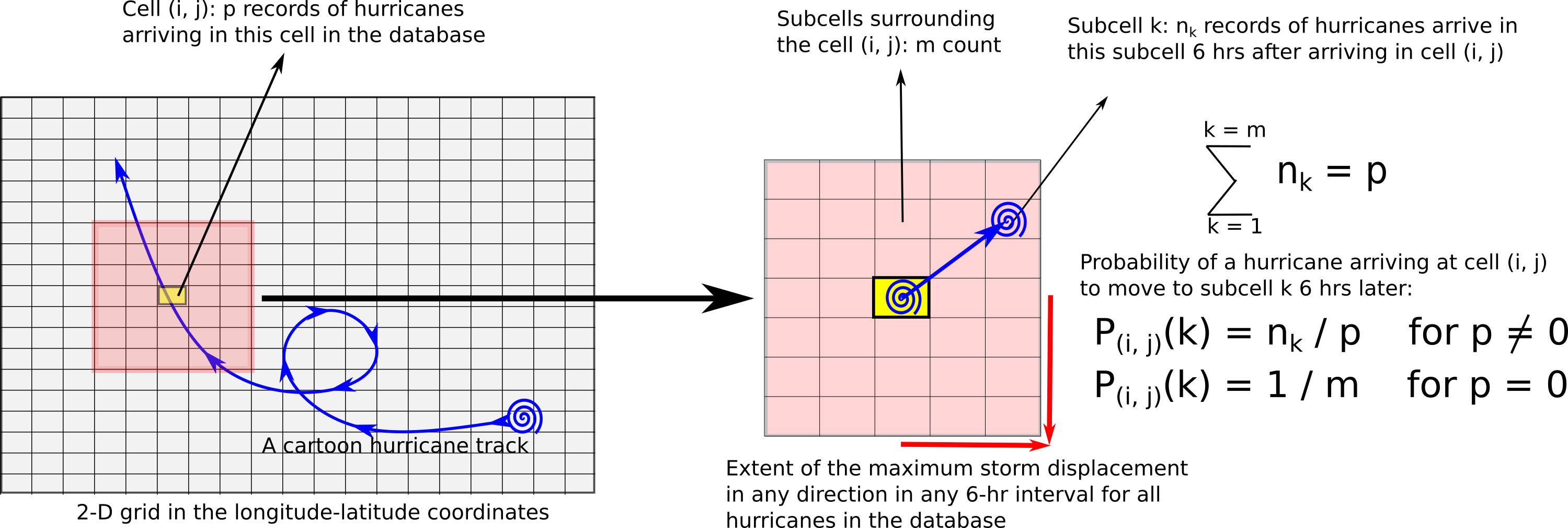}
        \caption{Schematic depicting the calculation of displacement probabilities associated with each cell in the grid domain.  This is a reproduction of Figure 1 from \cite{bose2021forecasting}.}
        \label{f1}
    \end{center}
\end{figure}


For trajectory models, the input features should contain information associated with trends of past storm motion \citep{emanuel2006statistical, moradi2016sparse}.
\citet{emanuel2006statistical} used a `transition matrix' to capture past trends and construct Markov chains. 
A similar approach is adopted here. 
We generate $6-hr$ displacement probabilities from DB-1 storms as input features of the DL trajectory models. 
A schematic depicting the calculation of displacement probabilities, taken from \citet{bose2021forecasting}, is shown in Fig. \ref{f1}. 
The $2-D$ domain bounded by the extents of $\lambda$ and $\phi$ (the extents are the maximum and minimum of each coordinate corresponding to any storm record in the considered database) are decomposed into rectangular cells. 
Consider all storms for which a record is contained in cell $(i, j)$ (colored yellow in Fig.~\ref{f1}). 
From cell $(i, j)$, the maximum numbers of cells traversed by a storm in the $\lambda-$ and $\phi-$ directions in any $6-hr$ interval are denoted by $m_\lambda$ and $m_\phi$, respectively. 
Therefore, associated with a cell $(i, j)$ there is a set of $m = (2m_\lambda + 1)(2m_\phi + 1)$ cells within which all $6-hr$ displacements of any historical storm are contained. 
Such a set is colored red in Fig.~\ref{f1}. 
Assume there are $p$ records of hurricanes that arrived at cell $(i, j)$. 
If, of the $p$ records, $n_k$ arrive at the $k^{th}$ cell, of $m$ cells, $\sum_1^m n_k = p$. 
The displacement probability of a storm in cell $(i, j)$ transitioning to the $k^{th}$ associated cell in the next 6 hours is calculated from the following expression.  

\begin{equation}
    p^{(i,j)}_k =   
    \begin{cases}
        \frac{n_k}{p},& \text{if } p\neq 0\\
        \frac{1}{m},& \text{if } p = 0
    \end{cases}
    \label{tP1}
\end{equation}

For a fine grid, the number of cells with $p=0$ would be larger than for a coarse grid. 
Further, for a finer grid 
$m$ would increase, and both $p$ and $n_k$ would decrease. 
In addition to an increased number of features ($m$) at each instant, displacement probabilities from a cell to adjacent cells would be noisier, which is not desirable for the prediction of new storms. 
On the other hand, if the grid is too coarse, trends reflected by the displacement probabilities could obscure the fine details of storm motion. 

Storm displacement probabilities were computed based on motions of DB-1 storms. 
In both the $\phi$ and $\lambda$ directions 61 grid points were used, resulting in the 2-D domain being decomposed into 3600 cells. 
It was found that for 1614 cells $p = 0$. 
Further, $(m_\lambda, m_\phi) $ = (6, 5) captured all $6-hr$ displacements. 
For any given cell, displacement probabilities were calculated for the $m = 13 \times 11$ = 143 associated cells. 
Therefore, for the DL translation models, the number of input features at each time record was 148 ($\phi$ and $\lambda$ coordinates, storm translation speed $V$ and direction $\theta$, wind speed $w_m$, and historical $6-hr$ displacement probabilities, $p_k$'s calculated at 143 cells associated to the cell containing $\lambda$ and $\phi$). 
The ML intensity models' input features include $p_c$ but not the displacement probabilities for 6 features in total. 
The input features for the different trajectory and intensity models considered herein are summarized in Tables \ref{t1} and \ref{t2}, respectively. 

\subsection{Trajectory models}
\label{sec:trajectory-models}

\begin{table}[!ht]
\caption{Input features to the ML and DL models for trajectory, and corresponding model outputs at $6-hr$ intervals. 
Here, $i$ denotes the $i^{th}$ time instant.}\label{t1}
\begin{tabular*}{\hsize}{@{\extracolsep\fill}lcccccc@{}}
\topline
Model (Abb.) & Architecture & Database & Input features & Predicted variables\\
\midline
\ Trajectory--DL & LSTM-RNN & DB-1 & $\lambda^{(i-1)}$, $\phi^{(i-1)}$, $V^{(i-1)}$, $\theta^{(i-1)}$, $w_m^{(i-1)}$, $p_k^{(i-1)}$ & $\Delta \lambda ^{(i-1)}$, $\Delta \phi^{(i-1)}$ \\
 (T-DL-1) & Many-To-Many-1 &  &  &  \\
\midline 
\ Trajectory--DL & LSTM-RNN & DB-1 & $\lambda^{(i-2)}$, $\phi^{(i-2)}$, $V^{(i-2)}$, $\theta^{(i-2)}$, $w_m^{(i-2)}$, $p_k^{(i-2)}$, & $\Delta \lambda ^{(i-1)}$, $\Delta \phi^{(i-1)}$, $\Delta \lambda ^{(i)}$, $\Delta \phi^{(i)}$ \\
(T-DL-2) & Many-To-Many-2 &  & $\lambda^{(i-1)}$, $\phi^{(i-1)}$, $V^{(i-1)}$, $\theta^{(i-1)}$, $w_m^{(i-1)}$, $p_k^{(i-1)}$ &  \\
\botline
\end{tabular*}
\end{table}

Two types of trajectory model were considered. 
One uses the RNN architecture with LSTM cells, invented by \citet{hochreiter1997long}, as the main recurrent unit. 
RNNs are utilized in DL to extract pattern and context from sequences and are used herein to advance a storm in time. 
The LSTM layers attend to the vanishing gradient issue in sequence problems. 
In an LSTM unit, relevant information from the past may be retained over long sequences via a cell state that passes through all LSTM layers. 
Early use of LSTMs in weather forecasting is reported by \citet{gomez2003local}. 
Except for the predicted quantities (see Table \ref{t1}), the LSTM-RNN models used herein are similar in all aspects to the models used in \cite{bose2021forecasting} to forecast the evolution of North Atlantic hurricanes from genesis. 

The second type of trajectory models are RFs. 
Compared to DL models, RF models require less data, and are therefore useful for the purpose of training a model that uses the input feature $p_c$, which is only available in DB-2 and DB-3, which contain far fewer storms than DB-1. 
An RF model with the large input feature set of the LSTM-RNN models, namely the 143 displacement probability features, is computationally prohibitive. 
Simulated storm databases produced by the two sets of trajectory models developed herein were statistically very similar, and therefore, results obtained using only the LSTM-RNN trajectory models are presented here. 

\citet{bose2021forecasting} showed that the Many-To-One type recurrent prediction architectures are susceptible to compounded error accumulation for long-term forecasting beyond the minimum recurrence period (i.e., 6 hours in the present case). 
The Many-To-Many prediction architectures were shown to reduce compounded error accumulation. 
Furthermore, the Many-To-Many-2 prediction architecture that takes in 2 time records as inputs and forecasts 2 time records in advance was shown to be the most accurate of all LSTM-RNN models for $6-hr$ short-term forecasts. 
Therefore, the Many-To-Many-2 prediction architecture is preferred for the trajectory models used herein in conjunction with the Many-To-Many-1 model which advances a storm 1-time step from its initial condition/genesis. 
In addition to the input and output layers, the bi-directional architecture was used, which consists of two bi-directional LSTM layers at either side of a repeat-vector layer for the Many-To-Many models.  
Each LSTM cell in each bi-directional layer had an input and output dimension of 64; so 128 neurons were used in each layer. 


Both LSTM-RNN and RF-ML models include $w_m$ as an input feature (input features to the RF-ML models also include $p_c$), which couples the trajectory models and intensity models. 
Previous works, with few exceptions, such as one of the trajectory models used by \citet{emanuel2006statistical}, did not attempt to couple the trajectory and intensity of a simulated storm. 
In reality, storm trajectories are dependent on the storm intensity and the ambient atmospheric conditions. 
Coupling of the trajectory and intensity models through the input features is a salient aspect of the present work.  


A set of input features at time steps $i-2$ and $i-1$ ($i-1$ at genesis) are fed into a trajectory model
to obtain $6-hr$ increments in $\lambda$ and $\phi$, $\Delta\lambda$ and $\Delta\phi$, for the next 2 time instants. 
The storm locations at time steps $i$ and $i+1$ are obtained by the following expressions, which involve the addition of error terms, $\epsilon_{\Delta \lambda, \Delta \phi}^{(+1, +2)}$ to the DL/ ML--model predictions.  
\begin{eqnarray}
\label{lam1}
\lambda^{(i)} = \lambda^{(i-1)} + \Delta \lambda^{(i-1)} + \epsilon_{\Delta \lambda}^{(+1)}	\\
\label{phi1}
\phi^{(i)} = \phi^{(i-1)} + \Delta \phi^{(i-1)} + \epsilon_{\Delta \phi}^{(+1)}	\\
\label{lam2}
\lambda^{(i+1)} = \lambda^{(i)} + \Delta \lambda^{(i)} + \epsilon_{\Delta \lambda}^{(+2)}	\\
\label{phi2}
\phi^{(i+1)} = \phi^{(i)} + \Delta \phi^{(i)} + \epsilon_{\Delta \phi}^{(+2)}
\end{eqnarray}
In these expressions, the error terms $\epsilon_{\Delta \lambda, \Delta \phi}^{(+1, +2)}$ are resampled from an error probability density function (PDF) obtained from the model predictions performed on test sequences not used for training.
The longitude and latitude errors are resampled jointly to maintain correlations.
The addition of the random error increases the likeness of simulated storms to real storms.  The predicted trajectories are unphysically smooth otherwise.
Similar error terms were included by other researchers for this purpose, such as, \citet{vickery2000simulation, vickery2009us, snaiki2020revisiting}. 

\subsection{Intensity models}
\label{sec:intensity-models}

Two sets of intensity models are created, one to predict $p_c$, and another to predict $w_m$. 
The input features to these models at different time steps, their abbreviated nomenclature, underlying model architecture, and the database used for model training/testing are listed in Table \ref{t2}. 
The underlying architecture is an RF because there are fewer storms with $p_c$ and $w_m$ provided for each time instant. 
Additionally, the distributions of $p_c$ and $w_m$ have heavy tails that are important for estimating extreme wind speeds and better accommodated by RF models. 
The RF models are built upon decision trees (DTs) as implied by the name \citep[see][for an example of a DT algorithm]{quinlan1986induction}. 
In an RF, the only tuned hyperparameter is the number of DTs, $n_{\rm estimators}$, ranging from 5 to 1000. 
In the present work, all input features were considered for splitting of the predictor variables. 
In this manner, each DT in an estimator was expanded up to its maximum depth, i.e., until the `leaves are pure', i.e., the hyperparameter $min$\textunderscore$samples$\textunderscore$leaf = 1$. 
The choice of $n_{\rm estimators}$ is based on the test/train split.  
During training, the whole training database is used and consequently tested on the test data set. 
The hyperparameter $n_{\rm estimators}$ is chosen that provides the minimum mean squared prediction error for the test set. 
 
The intensity models include the landfall status ($\zeta$) as a binary input feature at all input time steps (see Table \ref{t2}). 
Tropical storms dissipate on landfall because of the lack of latent heat from the ocean and of increased surface roughness. 
\citet{vickery2000simulation, vickery2009us, snaiki2020revisiting} modeled the storm-relative intensity by equations whose parameter coeffecients were functions of a storm's location, the sub-basin being traversed, intensity, heading etc. 
Upon landfall, the relative intensity of a storm was obtained using a dissipation model \citep{vickery1995wind}. 
The decay models while motivated by physical considerations, are essentially empirical. 
In a data-based approach, the signature characteristics of storm dissipation over land are inherent in the database and may be learned by the RF model. 
Distinguishing a storm's landfall status helps the ML models learn the dissipative charactersitics of a storm over land.  
The $\zeta$ input feature, a land fall status indicator, is found to be an important predictor for both $p_c$ and $w_m$. 
Its inclusion facilitates the use of the same prediction model for storm intensity over both land and sea; a separate storm dissipation model is no longer necessary.

\begin{table}[!ht]
\caption{Input features to the ML models for intensity, and corresponding model outputs at $6-hr$ intervals. 
Here, $i$ denotes the $i^{th}$ time instant.}\label{t2}
\begin{tabular*}{\hsize}{@{\extracolsep\fill}lcccccc@{}}
\topline
Model (Abb.) & Architecture & Database & Input features & Predicted variables\\
\midline
\ Intensity--ML: $p_c$ & RF-ML & DB-3 & $\lambda^{(i-1)}$, $\phi^{(i-1)}$, $V^{(i-1)}$, $\theta^{(i-1)}$, $w_m^{(i-1)}$, $p_c^{(i-1)}$, $\zeta^{(i-1)}$ & $p_c^{(i)}$ \\
(I-ML-1:$p_c$) & Many-To-Many-1 &  & $\lambda^{(i)}$, $\phi^{(i)}$, $\zeta^{(i)}$ & \\
\midline
\ Intensity--ML: $p_c$ & RF-ML & DB-3 & $\lambda^{(i-2)}$, $\phi^{(i-2)}$, $V^{(i-2)}$, $\theta^{(i-2)}$, $w_m^{(i-2)}$, $p_c^{(i-2)}$, $\zeta^{(i-2)}$ & $p_c^{(i)}$, $p_c^{(i+1)}$ \\
(I-ML-2:$p_c$) & Many-To-Many-2 & & $\lambda^{(i-1)}$, $\phi^{(i-1)}$, $V^{(i-1)}$, $\theta^{(i-1)}$, $w_m^{(i-1)}$, $p_c^{(i-1)}$, $\zeta^{(i-1)}$ & \\
 &  &  & $\lambda^{(i)}$, $\phi^{(i)}$, $\zeta^{(i)}$, $\lambda^{(i+1)}$, $\phi^{(i+1)}$, $\zeta^{(i+1)}$ & \\
\midline 
\ Intensity--ML: $w_m$ & RF-ML & DB-3 & $\lambda^{(i)}$, $\phi^{(i)}$, $V^{(i)}$, $\theta^{(i)}$, $p_c^{(i)}$, $\zeta^{(i)}$ & $w_m^{(i)}$ \\
(I-ML-1:$w_m$) & Many-To-Many-1 &  &  & \\
 \midline 
\ Intensity--ML: $w_m$ & RF-ML & DB-3 & $\lambda^{(i)}$, $\phi^{(i)}$, $V^{(i)}$, $\theta^{(i)}$, $p_c^{(i)}$, $\zeta^{(i)}$ & $w_m^{(i)}$, $w_m^{(i+1)}$ \\
(I-ML-2:$w_m$) & Many-To-Many-2 &  & $\lambda^{(i+1)}$, $\phi^{(i+1)}$, $V^{(i+1)}$, $\theta^{(i+1)}$, $p_c^{(i+1)}$, $\zeta^{(i+1)}$ & \\
\botline
\end{tabular*}
\end{table}

Modelling of storm intensity based on a storm's location is critical as the intensity of a storm is highly dependent on ambient conditions such as the sea surface temperature, salinity, etc. 
These conditions vary among sub-basins/regions, and therefore, significant variability of storm intensity is expected between different regions. 
This is difficult to capture using a single model, but training a different model for each region would require training more models than is practical. 
Instead, three sets of intensity models are trained based on the magnitude of $w_m$. 
The first set is applied if at all input time instants $w_m$ remains below hurricane wind speeds, i.e., $w_m < 33$ $ms^{-1}$; the second set of models is applied if a storm attains up to category-3 hurricane wind speeds, i.e., at any of the input time instants $33\; ms^{-1} \le w_m < 58\; ms^{-1}$; the final set of models is applied in case a storm attains category-4 hurricane wind speeds or above, i.e., $w_m \ge 58$ $ms^{-1}$. 
The model names, databases used for training, and $w_m$ ranges are detailed in Table \ref{t3}. 
In this flexible strategy, each segment of the distribution of $w_m$ is modeled separately by manipulation of the training database, DB-3; complexity associated with the use of zonal/regional intensity models may be avoided. 
On the other hand, the correct spatial distribution of high wind speed records is retained. 
Most importantly, the high wind speeds in category 4 and 5 hurricanes may be modeled accurately instead of the being "smoothed" by the much larger number of low intensity storms. 

\begin{table}[!ht]
\caption{Models trained and applied based on the magnitude of $w_m$.}\label{t3}
\begin{tabular*}{\hsize}{@{\extracolsep\fill}lccccc@{}}
\topline
Models & Database & Criterion of minimum $w_m$ for training & Application criterion \\
\midline
\ $I-ML-2:p_c$, $I-ML-2:w_m$ & DB-3 & $w_m \ge 0$ $m/s$ & $w_m < 33$ $m/s$ &\\
\midline 
\ $I-ML-2:p_c$, $I-ML-2:w_m$ & DB-3 & $w_m \ge 25$ $m/s$ & 33 $m/s$ $\le w_m < 58$ $m/s$ &\\
\midline 
\ $I-ML-2:p_c$, $I-ML-2:w_m$ & DB-3 & $w_m \ge 50$ $m/s$ & $w_m \ge 58$ $m/s$ &\\
\botline
\end{tabular*}
\end{table}

For predictions of $p_c$ and $w_m$, an error term resampled from the PDFs of prediction errors calculated on test data sequences is added to the model predicted quantity using the set of equations
 
 \begin{eqnarray}
 \label{gam1}
 \gamma^{(i)} = \gamma_m^{(i)} + \epsilon_{\gamma}^{(+1)}	\\
 \label{gam2}
 \gamma^{(i+1)} = \gamma_m^{(i+1)} + \epsilon_{\gamma}^{(+2)},
 \end{eqnarray}
 where $\gamma$ is a prediction of either $p_c$ or $w_m$, and $\gamma_m$ is the prediction directly from the RF model.  The superscripts for the error terms $\epsilon_{\gamma}^{(+1)}$ and $\epsilon_{\gamma}^{(+2)}$ indicate the prediction step corresponding to the $I-ML-1:\gamma$ and $I-ML-2:\gamma$ models from Table \ref{t3}, respectively. 
 The error PDFs are constructed using the Gaussian Kernel Density Estimator (KDE) of the model prediction errors computed on test sequences. 
Joint resampling of the error $\epsilon_{\gamma}^{(+1,+2)}$ for the $I-ML-2:\gamma$ models ensures maintenance of correlations. 
 Addition of these error terms is necessary to incorporate variability in model predictions. Otherwise, for a given set of input features, any of the intensity models would always output the same $\gamma_m$.
 In reality, based on the ambient conditions, a storm may evolve differently starting from the same initial conditions due to the chaotic nature of the governing physics of storm evolution. 
 
\subsection{Storm origination}

The trajectory and intensity models described to this point require a complete set of input features to make their first prediction.  The initial input features, with the exception of $p_c$, are simulated from a joint probability distribution.  The probability distribution was constructed using the initial conditions from all storms with more than six time records and a positive initial speed, $V>0$.  Out of the 1893 storms in HURDAT2, 1719 were used.  The joint distribution was constructed by first fitting a Johnson $S_U$ distribution \citep[see for example][chapter 12]{johnson1994} to the histogram of each input feature, i.e., the marginal distribution of each input feature.  A Gaussian copula is used to combine the univariate marginal probability distributions into a joint probability distribution.  Call it $F_I$.  A detailed treatment of copulas may be found in \cite{nelsen2007introduction}.  The covariance matrix for the Gaussian copula was taken to be the observed correlation matrix of the input features.  The initial input features for a simulated hurricane are sampled from $F_I$, but after sampling, a rejection step \citep[see for example 10.3 of][]{gelman2013bayesian} is performed to ensure that the sampled proportion of storms from each of the five sub-basins matches the observed proportion.

An initial value for $p_c$ must also be generated.  Of the 1719 storms used to construct $F_I$, 603 have $p_c$ recorded.  For those storms, multiple linear regression \citep[see for example][]{neter1996applied} is used to predict $p_c$ based on the values of the other input features.  The root mean square prediction error of the multiple linear regression is less than 1\%.  

The prescription to generate an initial value for a simulated hurricane is to sample from $F_{\rm I}$ first.  Then perform a rejection step to ensure agreement with zone proportions. Last, use the sampled input features to predict $p_c$, yielding a complete initial condition.

One more issue remains.  To simulate one full year of hurricanes, the number of hurricanes to simulate in that year, which is random, must be selected.  This is done using a negative binomial distribution \citep{johnson2005} which is fitted to counts of hurricanes per year in HURDAT2 since 1975.

\begin{figure}
    \centering
    \includegraphics[width=6in]{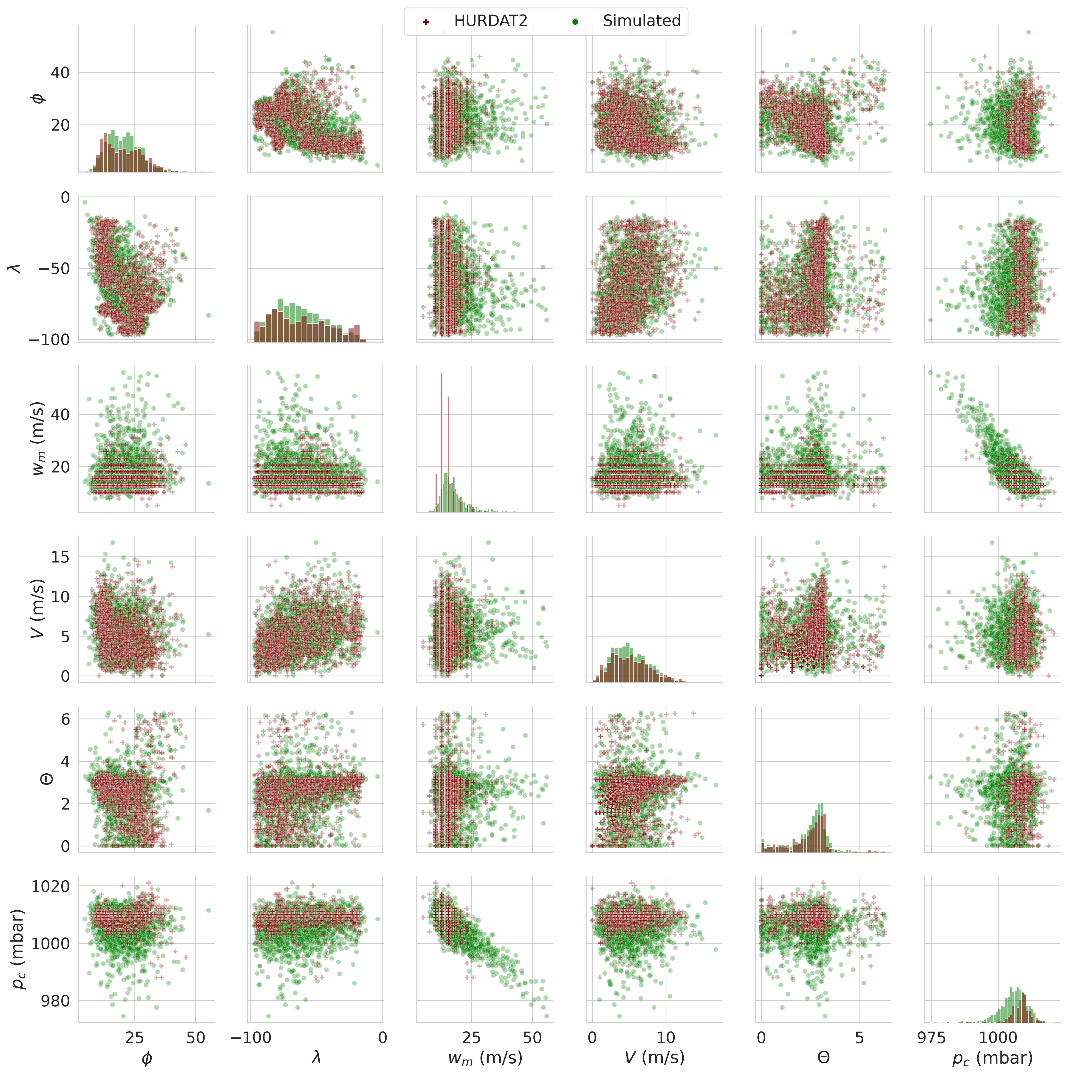}
    \caption{Initial storm conditions from HURDAT2 (plus signs) and simulated initial storm conditions (points).}
    \label{fig:hurdat_vs_sim_init}
\end{figure}

Figure \ref{fig:hurdat_vs_sim_init} shows a comparison of initial conditions from HURDAT2 (crosses) to simulated (dots).  Generally, we find that the simulated initial conditions match those found in HURDAT2 well.  In all cases, a convex hull around the simulated initial conditions would envelop the corresponding convex hull for the initial conditions from HURDAT2.  Thus, the simulation approach is successfully generating all important initial condition combinations, according to HURDAT2.


\subsection{Simulation methodology}

\begin{figure}[!ht]
    \begin{center}
        \includegraphics[trim=0cm 0cm 0cm 0cm,clip,width=0.95\textwidth]{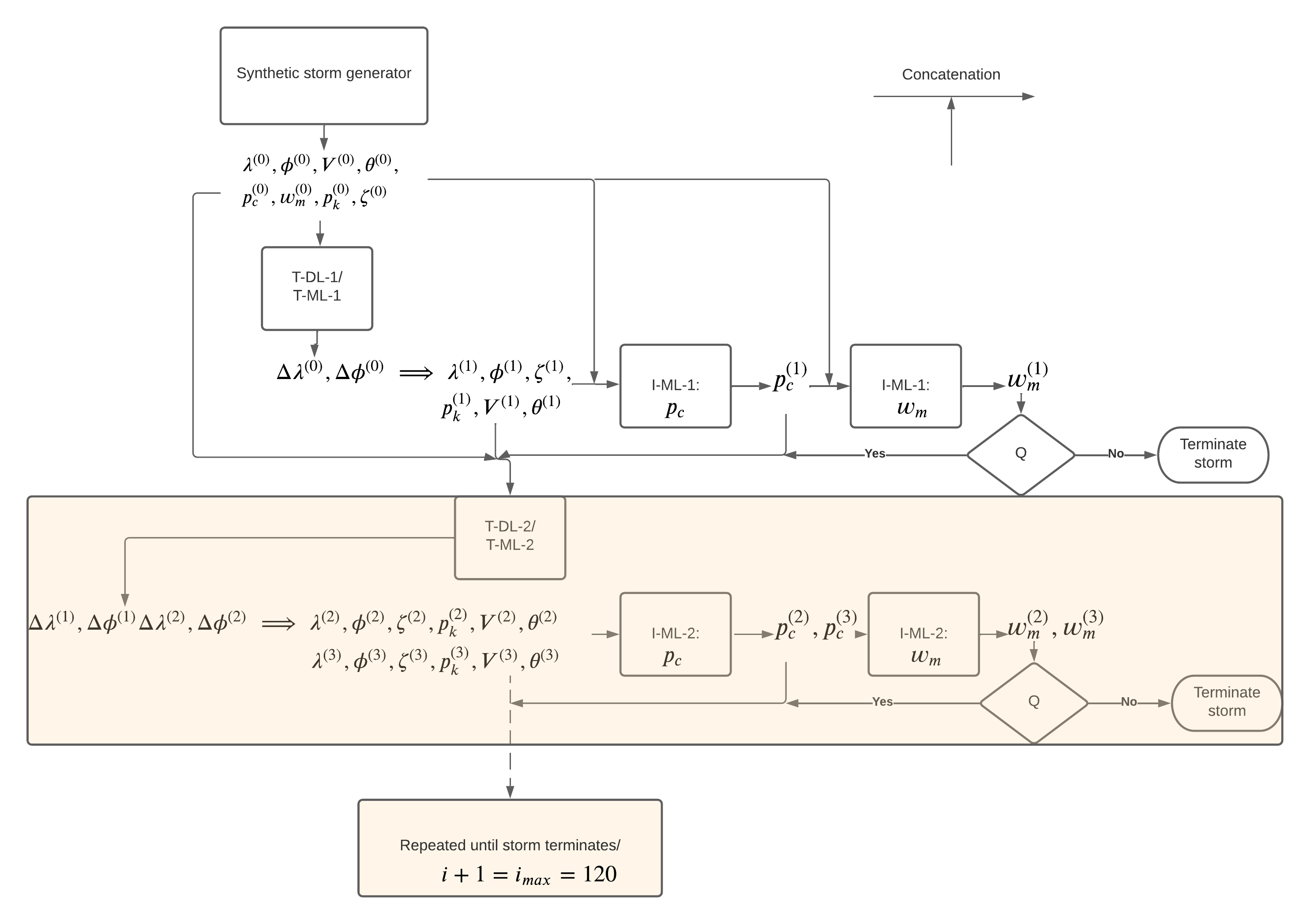}
        \caption{Schematic depicting the track generation method using the models described in Tables \ref{t1} and \ref{t2}.}
        \label{ftgm}
    \end{center}
\end{figure}

A flow chart of the simulation procedure to generate a single storm using the trajectory and intensity models is presented in Fig.~\ref{ftgm}. 
The synthetic storm genesis model provides an initial condition (time step $i = 0$ in Fig.~\ref{ftgm}) for a storm. 
The input features at $i = 0$ are fed to the trajectory model $T-DL-1$/ $T-ML-1$, which predicts the $6-hr$ increments in $\lambda$ and $\phi$ at $i=0$, i.e., $\Delta \lambda^{(0)}$ and $\Delta \phi^{(0)}$. 
The increments provide the storm location at $i=1$ using Eqns. \ref{lam1} and \ref{phi1}.
Next, the the $I-ML-1:p_c$ model and Eqn. \ref{gam1} are used to obtain $p_c^{(1)}$ a necessary input to the $I-ML-1:w_m$ model.
The $I-ML-1:w_m$ model and Eqn. \ref{gam1} may then be used to predict $w_m^{(1)}$.
This completes the execution of the prediction time step $i=1$. 
Once the input features at more than one time step are available, the trajectory model $T-DL-2$/ $T-ML-2$ in conjunction with the intensity models $I-ML-2:p_c$ and $I-ML-2:w_m$ are used. 
The execution involves the same aforementioned procedure; however, instead of one step at a time, predicted quantities are obtained two steps at a time. 
For example, if all storm features are available at $i-2$ and $i-1$, the $T-DL-2$/ $T-ML-2$ model outputs the next two $6-hr$ increments $\Delta \lambda^{(i-1, i)}$ and $\Delta \phi^{(i-1, i)}$. 
Using Eqns. \ref{lam1}--\ref{phi2}, the storm locations are obtained as $\lambda^{(i, i+1)}$ and $\phi^{(i, i+1)}$. Then, $p_c^{(i, i+1)}$ and $w_m^{(i, i+1)}$ are obtained sequentially from $I-ML-2:p_c$ and $I-ML-2:w_m$, respectively, and Eqns. \ref{gam1} and \ref{gam2}.
The algorithm is repeatedly applied (colored box in Fig. \ref{ftgm}) utilizing the 2-time step DL-2/ ML-2 models until the storm is terminated. 

After completion of each prediction time step, several checks are performed denoted as $Q$ in Fig. \ref{ftgm}. 
The conditions that must be satisfied at each time step $i$ for the procedure to advance to the next prediction step are listed below. The storm is otherwise terminated. 
\begin{itemize}
\item $w_m^{(i)} \ge w_m^{\rm critical}$
\item $w_m^{(i)} > V^{(i)}$
\item $102.5^\circ W \ge \lambda \ge 10^\circ W$
\item $50^\circ N \ge \phi \ge 0^\circ N$
\item $i+1 < i_{\rm max} = 120$
\end{itemize}

The first criterion ensures termination of a storm with low $w_m < w_m^{critical}$, i.e., the critical $w_m = 8 m/s$. 
The other constraint on $w_m$ is based on the physical consideration that for a Rankine vortex model to be valid, $w_m$ must be greater than the translation speed of the storm. 
The next two criteria ensure that the calculations are performed in the computational domain bounding box defined as $\phi \in [0^\circ N, 50^\circ N]$ and $\lambda \in [10^\circ W, 102.5^\circ W]$. 
The last criterion limits the life span of a storm. 
A storm may have a maximum of $i_{\rm max}=120$ time records, i.e., the maximum life span of a storm is $6i_{\rm max} = 720$ hrs.

\section{Results: Efficacy of the synthetic storm simulation model}\label{sec3}

Synthetic storms are generated in sets of $100-yr$ periods. 
In total five synthetic storm databases are simulated and compared with the storms in HURDAT2 since 1920 (1302 storms). 
The initial/ genesis conditions for the simulated storms are obtained as described in Sec. \ref{sec2}, which is based on storms occurring after 1975. 
Due to enhanced storm detection and tracking capabilities, and possibly also to the effect of global warming, the average number of storms listed per year in the HURDAT2 database has increased since the beginning of the satellite era in the 1970's; therefore, the average number of storms generated by the synthetic storm generator for a $100-yr$ period is generally larger than 1302. 

\subsection{Efficacy of the trajectory models}

\begin{figure}[!ht]
    \begin{center}
        \includegraphics[trim=0cm 0cm 0.0cm 0cm,clip,width=0.75\textwidth]{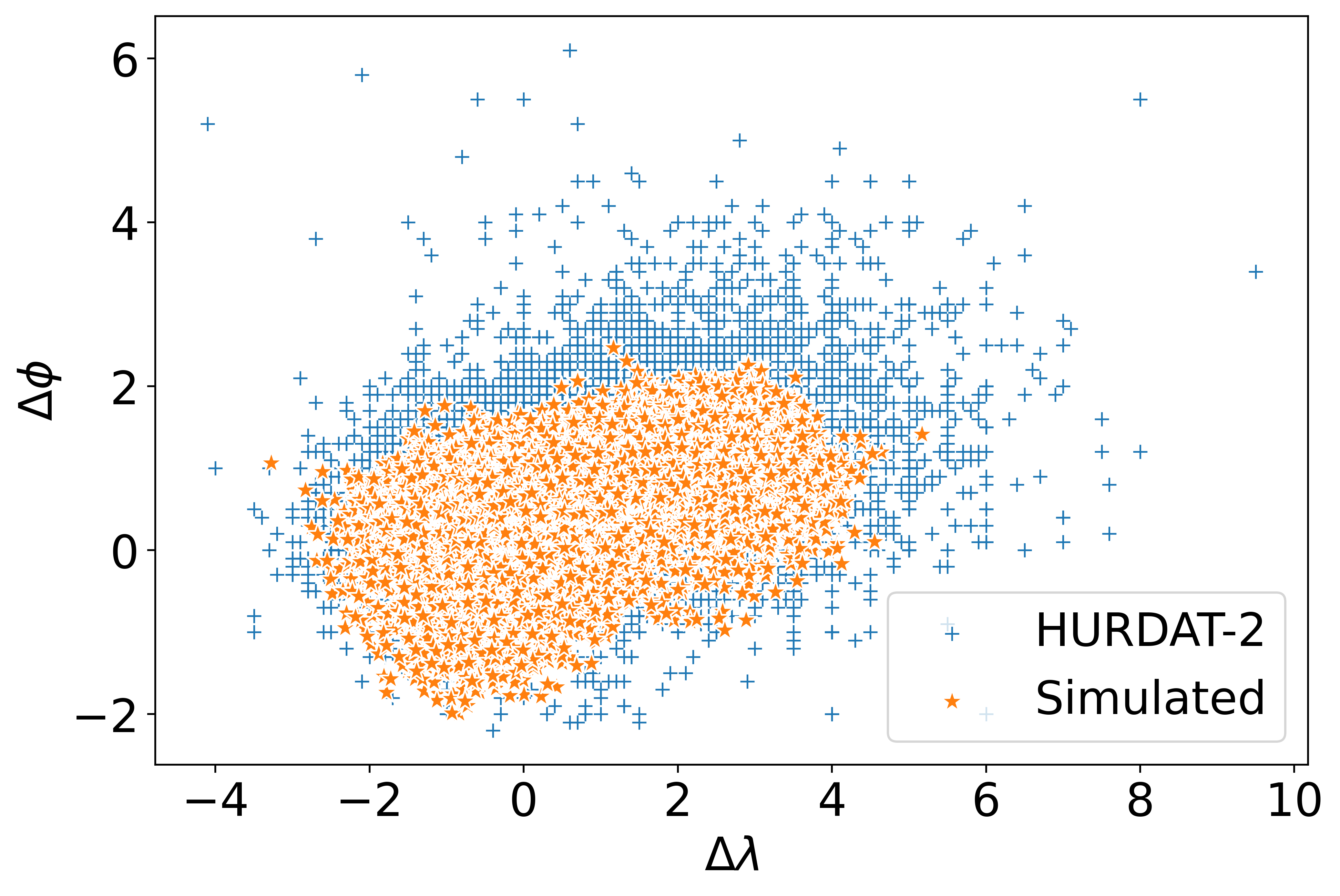}
        \caption{Joint scatterplot of $6-hr$ increments in latitude ($\Delta \phi$) and longitude ($\Delta \lambda$) for generated synthetic storms for a $100-yr$ period compared with storms from the HURDAT2 database since 1920.
        Records within the domain bounding box given by $\phi \in [0^\circ N, 50^\circ N]$ and $\lambda \in [10^\circ W, 102.5^\circ W]$ are used.}
        \label{f3}
    \end{center}
\end{figure}

\begin{figure}[!ht]
    \begin{center}
        \includegraphics[trim=0cm 0cm 0.0cm 0cm,clip,width=0.45\textwidth]{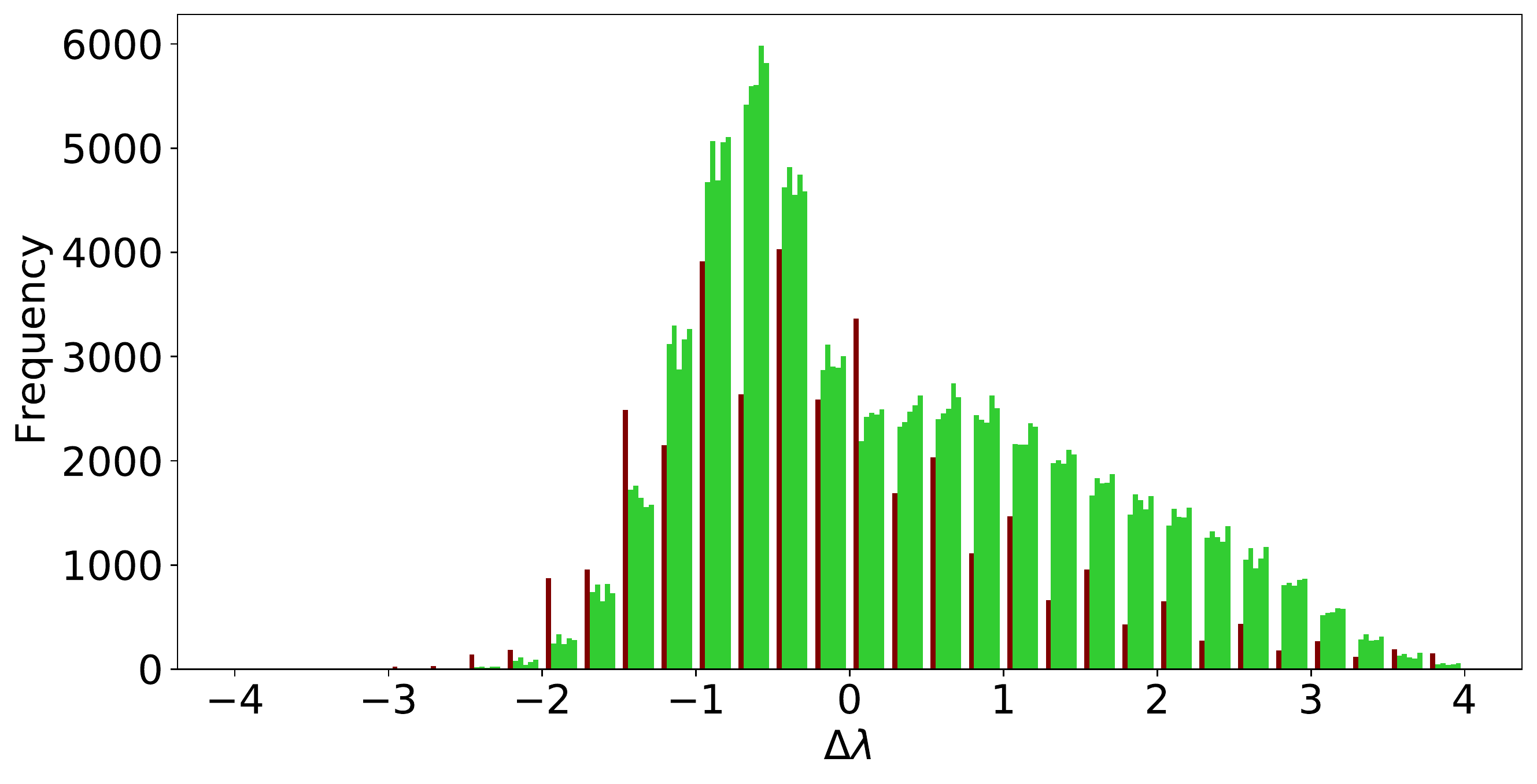}    
        \includegraphics[trim=0cm 0cm 0.0cm 0cm,clip,width=0.45\textwidth]{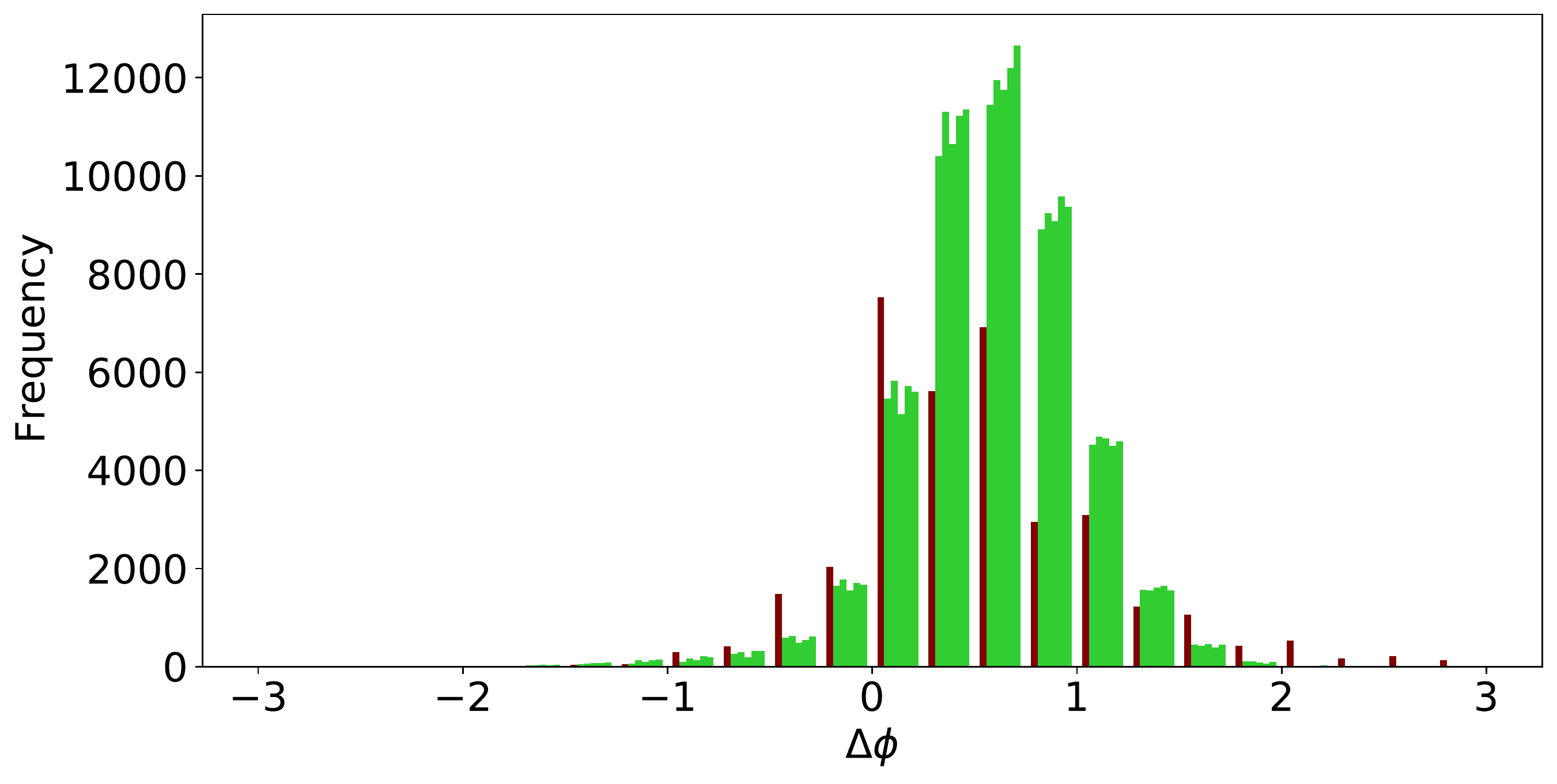}
        \caption{Frequency distributions of: $6-hr$ increments, $\Delta \lambda$ (left) and $\Delta \phi$ (right) for five simulated storm databases each computed for a $100-yr$ period (green) compared with storms from the HURDAT2 database since 1920 (maroon). 
        Records within the domain bounding box given by $\phi \in [0^\circ N, 50^\circ N]$ and $\lambda \in [10^\circ W, 102.5^\circ W]$ are used.}
        \label{f4}
    \end{center}
\end{figure}

The joint distribution of the $6-hr$ increments in $\phi$ (i.e., $\Delta \phi$) and $\lambda$ (i.e., $\Delta \lambda$) for the simulated storms over a $100-yr$ period (one of the five simulated storm databases) is compared to its HURDAT2 counterpart in Fig. \ref{f3}. 
The individual frequency distributions of $\Delta \phi$ and $\Delta \lambda$ are also shown for all five simulated storm databases and compared with the same HURDAT2 storm records in Fig. \ref{f4}. 
As the simulated storms were generated in the computational domain bounded by $\phi \in [0^\circ N, 50^\circ N]$ and $\lambda \in [10^\circ W, 102.5^\circ W]$, the HURDAT2 records since 1920 with storm positions within these limits are used for comparison. 
Figure~\ref{f3} suggests that $6-hr$ storm movements in HURDAT2 are predominantly in the north-east direction. 
The simulated storms generally agree with that trend, and the storm motions in other quadrants, i.e., $6-hr$ storm motions towards the south-west, north-west or south-east are also captured well by the simulation. 
The most extreme storm movements, i.e., $\Delta\phi>2$ and $\Delta\lambda>4.5$ are not well captured by the trajectory models. 
This is clear from Figure \ref{f3}, and is attributable to the fact that, like all regression models, DL and ML models may be thought of as smoothers, which is the primary reason for the added error terms in Eqns. \ref{lam1}--\ref{gam2}. 

Figure~\ref{f3} only provides a qualitative view, which is complemented by the quantitative view presented in Fig. \ref{f4}. 
 The frequency distributions show that both the left and right tails of $\Delta \phi$ and $\Delta \lambda$ from HURDAT2 are very thin and are underpredicted by the models. 
 Both distributions for the simulated storms are slightly shifted towards the right. 
The frequency in the mid-range for both quantities is over-predicted. 
This could be due to two reasons. 
First, as previously mentioned, the synthetic storm generator generates a larger number of storms in a chosen time interval. 
Second, the storm trajectories could be in general longer for the simulated storms. 
This is investigated later. 

\begin{figure}[!ht]
    \begin{center}
        \includegraphics[trim=0cm 0cm 0.0cm 0cm,clip,width=0.45\textwidth]{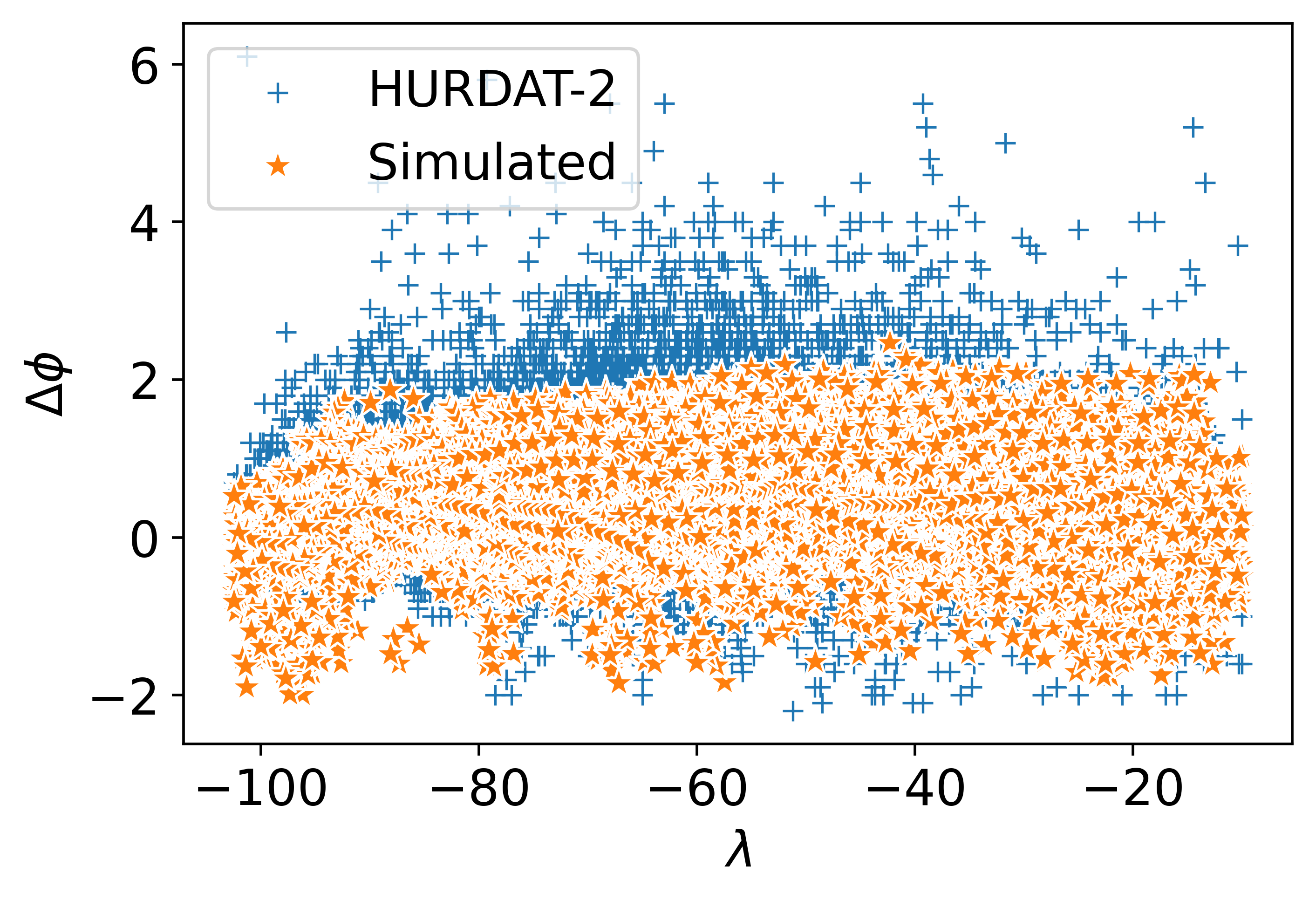}    
        \includegraphics[trim=0cm 0cm 0.0cm 0cm,clip,width=0.45\textwidth]{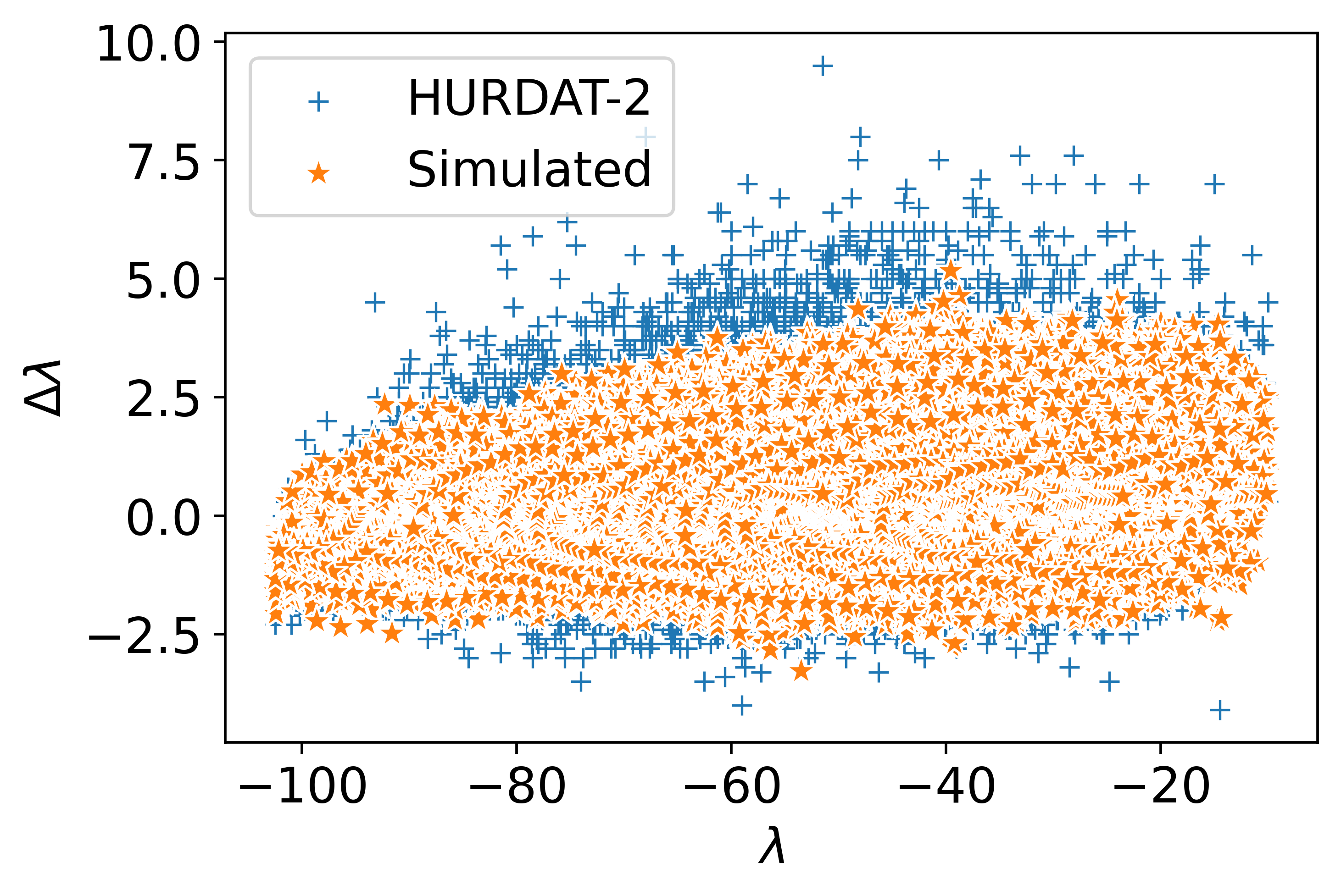}
        \includegraphics[trim=0cm 0cm 0.0cm 0cm,clip,width=0.45\textwidth]{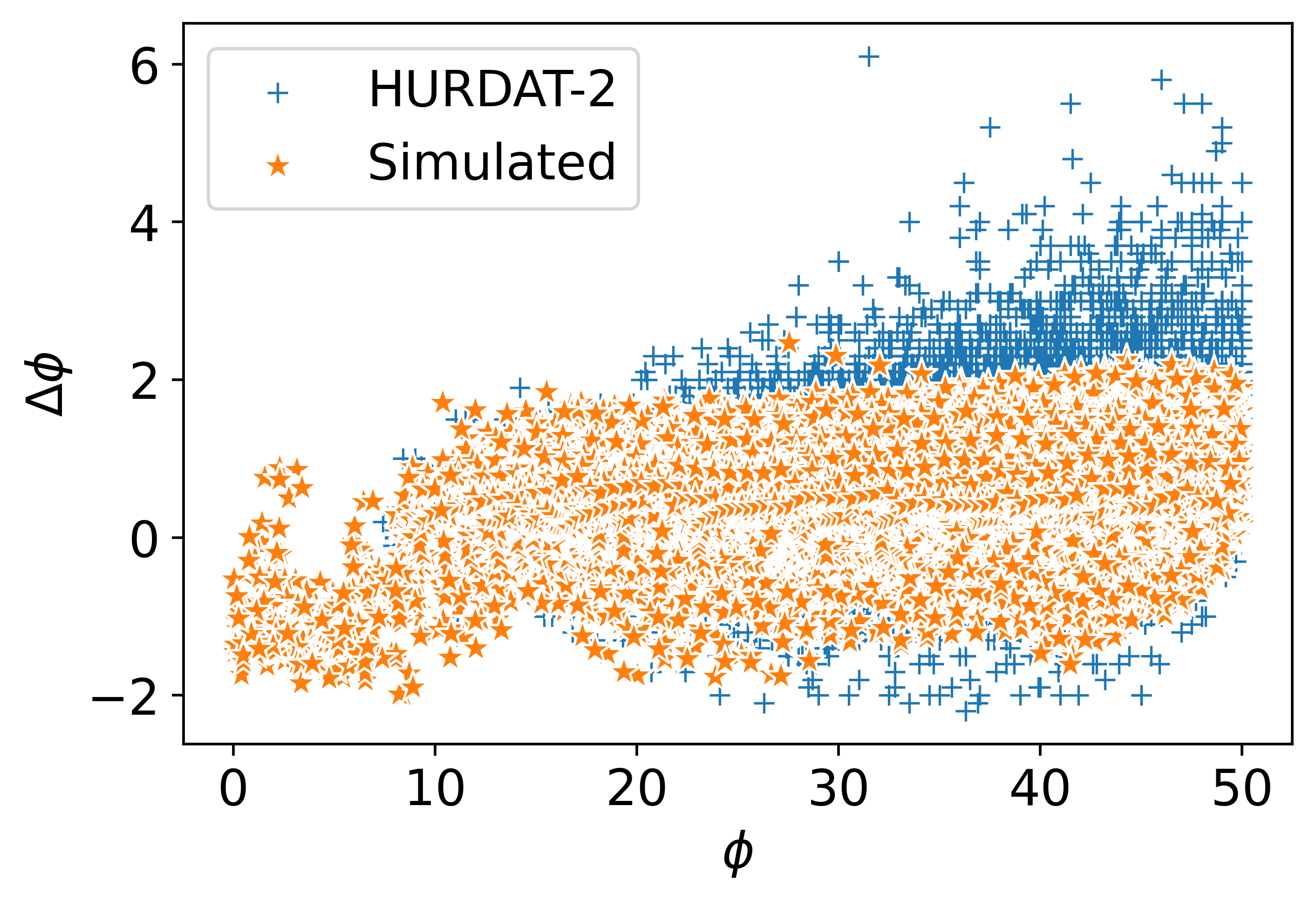}    
        \includegraphics[trim=0cm 0cm 0.0cm 0cm,clip,width=0.45\textwidth]{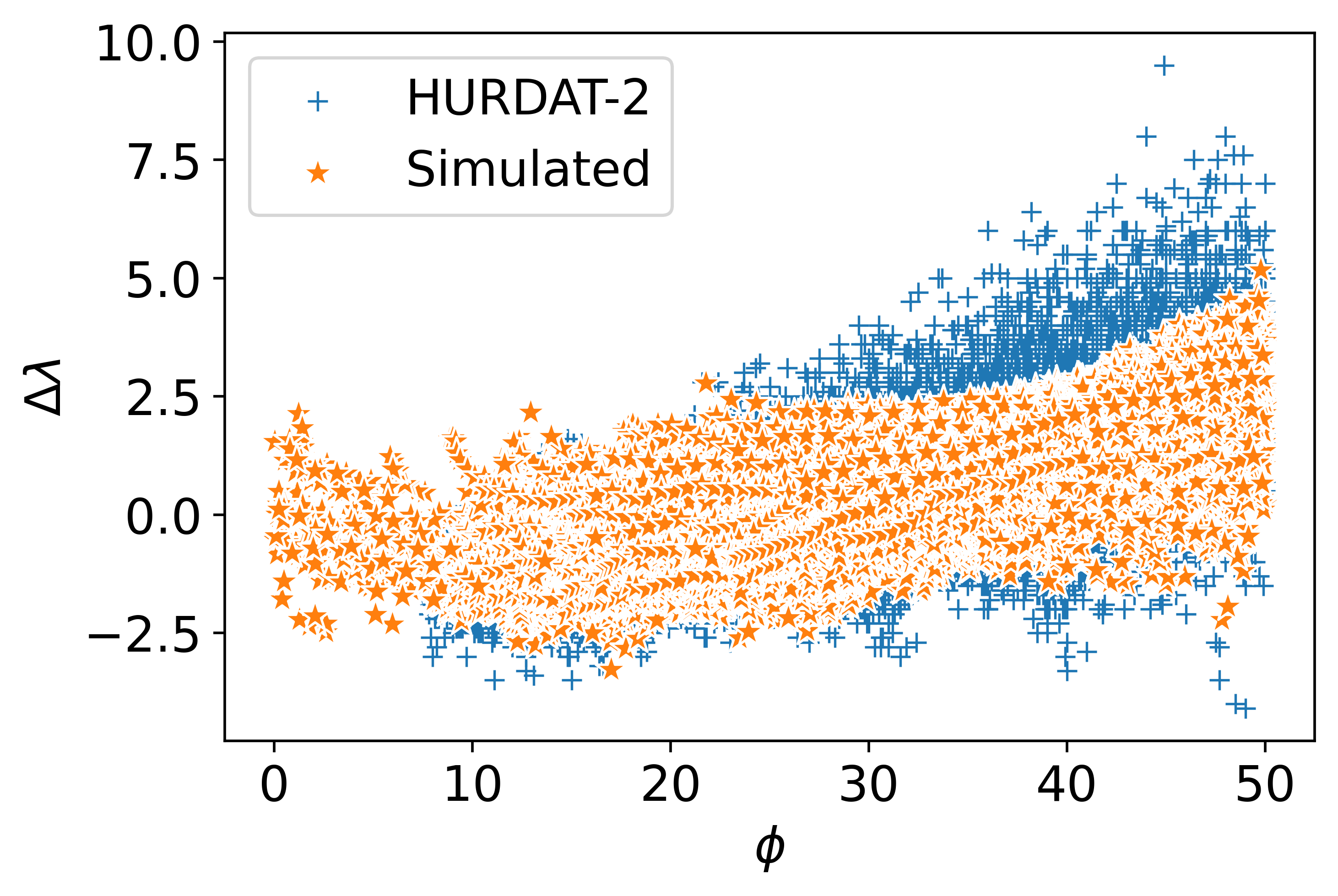}
        \caption{Joint scatterplots showing $6-hr$ storm motion as a function of storm's location: $\Delta \phi$ vs. $\lambda$ (top left), $\Delta \lambda$ vs. $\lambda$ (top right), $\Delta \phi$ vs. $\phi$ (bottom left) and $\Delta \lambda$ vs. $\phi$ (bottom right) for generated synthetic storms for a $100-yr$ period compared with storms from the HURDAT2 database since 1920.}
        \label{f5}
    \end{center}
\end{figure}

It is also informative to plot the combinations $\{\Delta\phi, \Delta\lambda\}\times\{\phi, \lambda\}$, which are found in Fig. \ref{f5}.
The scatterplots show that the right tails of the HURDAT2 distributions are not captured by the trajectory models. 
This again is expected on account of the inherent smoothing associated with the use of the LSTM-RNN models. 
However, the relevant ranges of these increments are adequately represented for the simulated storms. 

\begin{figure}[!ht]
    \begin{center}
        \includegraphics[trim=0cm 0cm 0.0cm 0cm,clip,width=0.75\textwidth]{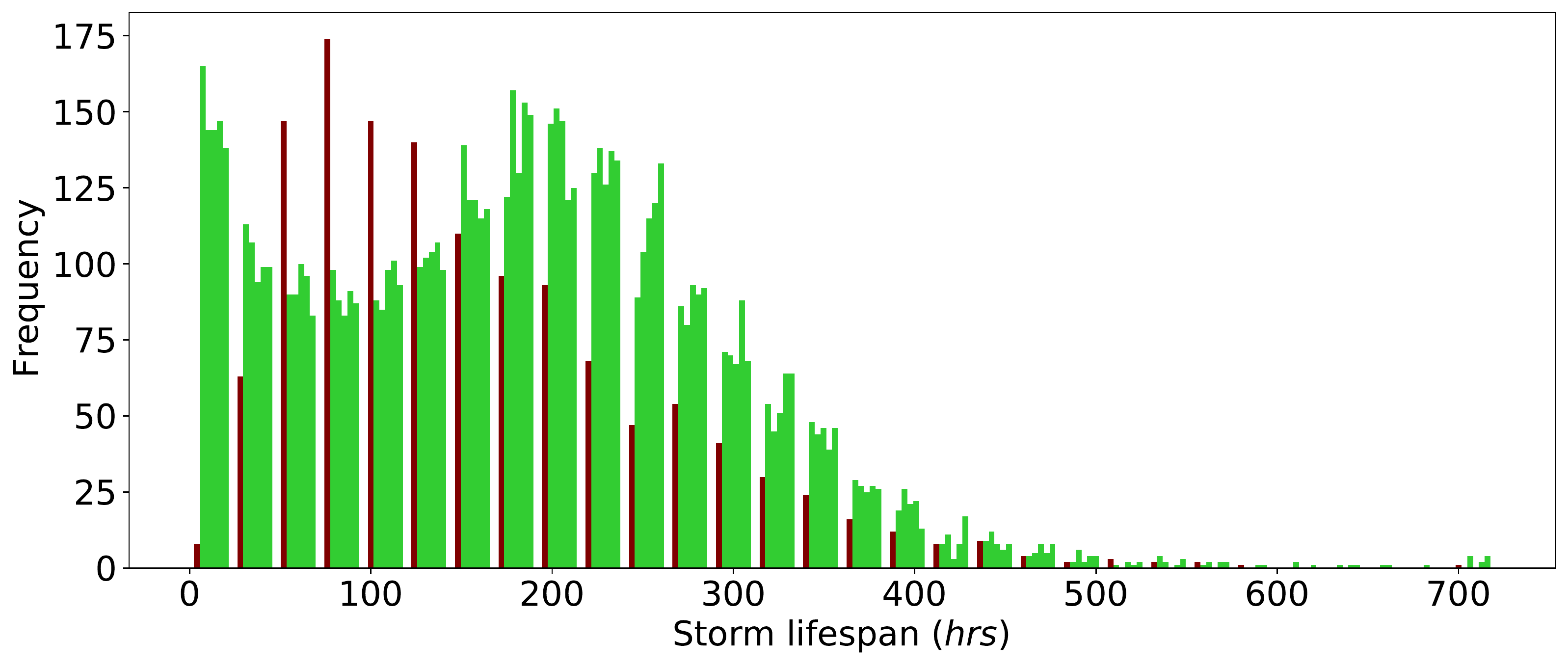} 
        \includegraphics[trim=0cm 0cm 0.0cm 0cm,clip,width=0.75\textwidth]{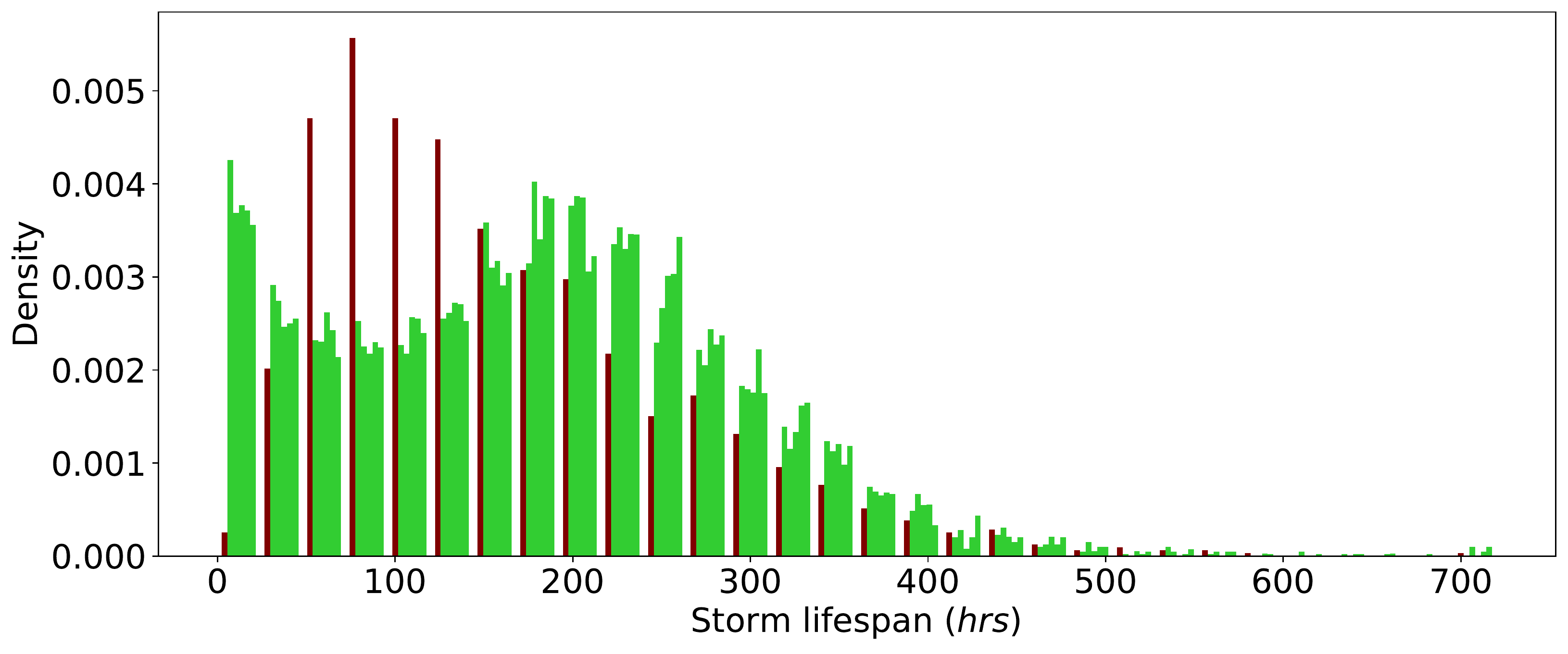} 
        \caption{Frequency distribution (top) and the corresponding PDF (bottom) of lifespan of storms in the five simulated storm databases, each simulated for a $100-yr$ period (green) compared with that from the HURDAT2 database since 1920 (maroon). 
        There are 1302 storms since 1920 in the HURDAT2 database compared to an average of 1619 storms in each simulated storm database over the same time interval.}
        \label{f6}
    \end{center}
\end{figure}

The frequency distributions and the corresponding Probability Density Functions (PDFs) of the lifespan of storms in hours for the five simulated storm databases are compared with their HURDAT2 counterpart in the top and bottom frames of Fig. \ref{f6}, respectively. 
The PDFs are shown to account for the larger number of storms in the simulated storm databases. 
It is noted that the lifespan of a simulated storm is dependent on the efficacy of the intensity models. 
The frequency distributions are in good agreement only in the long lifespan range above 400 $hrs$ (bottom frame of Fig. \ref{f6}). 
HURDAT2 has a larger proportion short-lifespan storms (lifespan between 48 and 144 hours) than the simulated storm databases. 
On the other hand, the simulated databases have a larger proportion of medium lifespan range storms (between 168 hrs. and 400 hrs.) and very short lifespan storms (lifespan below 48 hrs.). 
The discrepancies are possibly due to the use of DB-3 for storm intensity modelling. 
Overemphasis on storms that attain larger values of $w_m$ may result in storms of longer duration. 
Also, underemphasis on low values of $w_m$ means that storms with low $w_m$ die out sooner than HURDAT2 storms. 
This results in the large overshoot in the very low lifespan range. 
In spite of the aforementioned issue related to the use of DB-3, storm intensities are well represented both globally (see the next subsection) and locally (Sec. \ref{sec4}). 
Therefore, it is likely that the overestimation in the frequency distributions of $\Delta \phi$ and $\Delta \lambda$ in Fig. \ref{f4} is due to the larger number of storms in the simulated databases, and to overemphasising of high $w_m$ storms in the model training database for the intensity models. 

\begin{figure}[!ht]
    \begin{center}
        \includegraphics[trim=0cm 0cm 0.0cm 0cm,clip,width=0.45\textwidth]{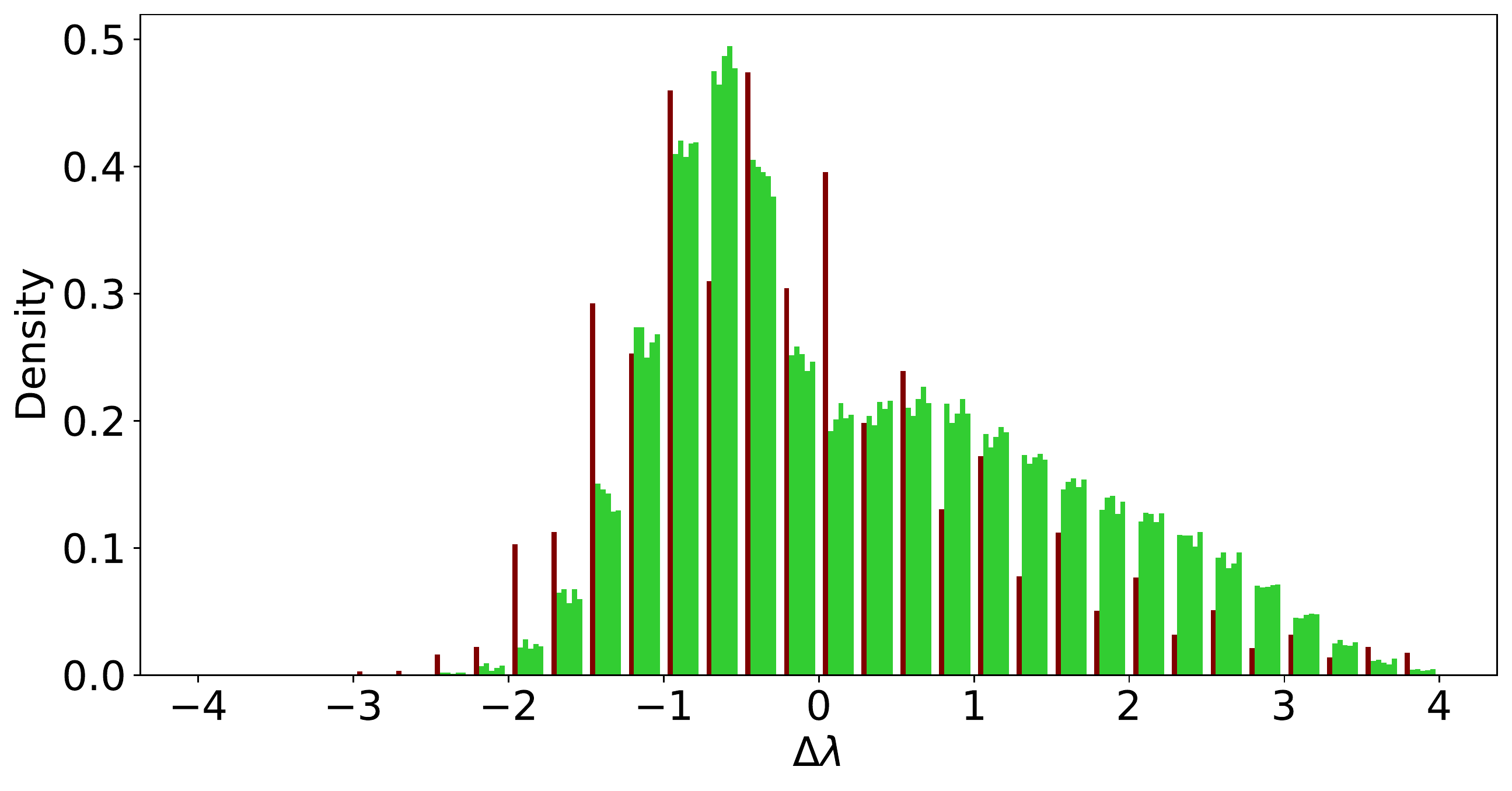}    
        \includegraphics[trim=0cm 0cm 0.0cm 0cm,clip,width=0.45\textwidth]{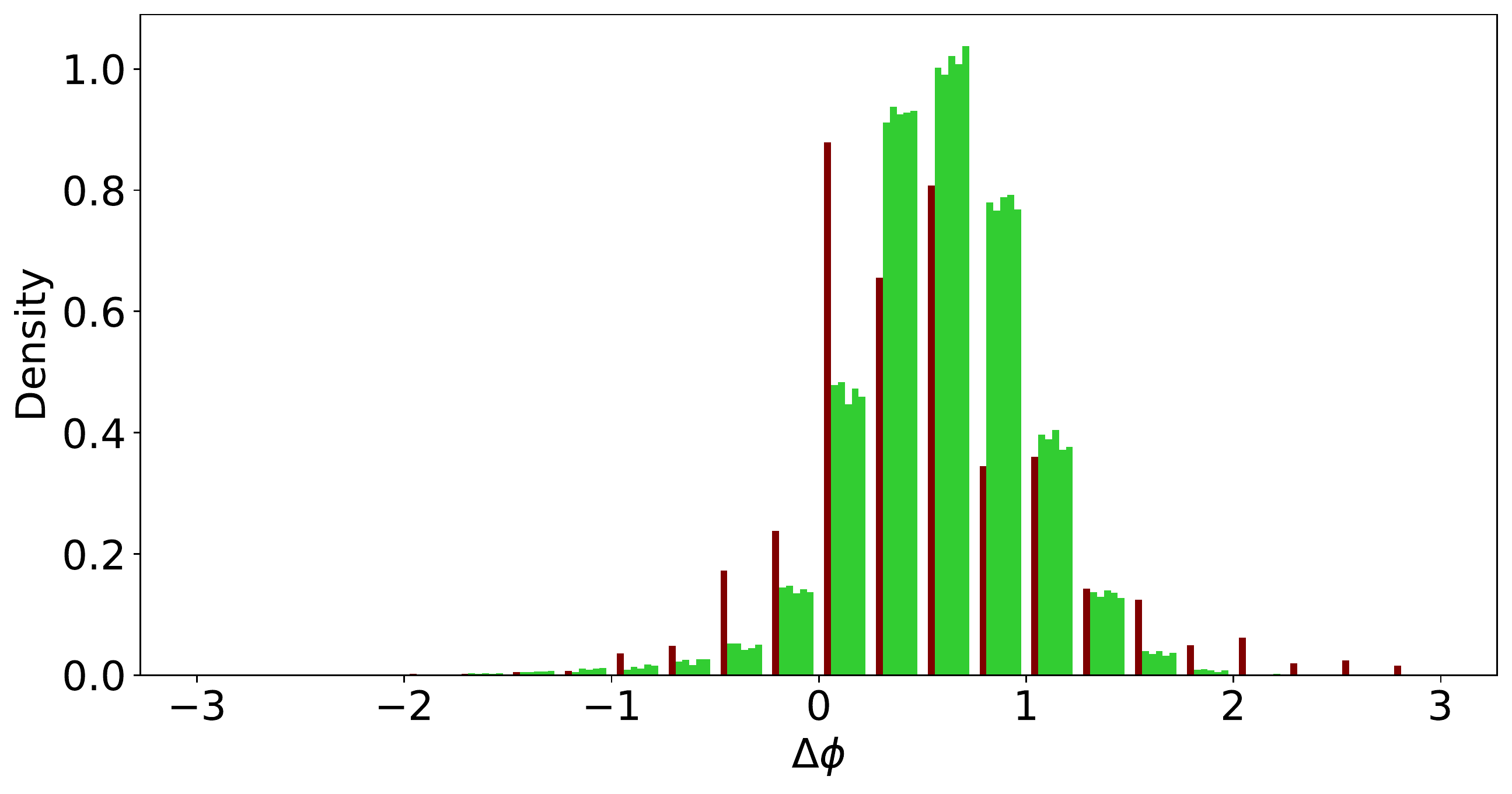}
        \caption{PDFs of: $6-hr$ increments $\Delta \lambda$ (left) and $\Delta \phi$ (right) for five simulated storm databases each computed for a $100-yr$ period (green) compared with storms from the HURDAT2 database since 1920 (maroon). 
        Records within the domain bounding box given by $\phi \in [0^\circ N, 50^\circ N]$ and $\lambda \in [10^\circ W, 102.5^\circ W]$ are used.}
        \label{f4a}
    \end{center}
\end{figure}

In order to account for the larger number of storms in the five simulated storm databases, the PDFs of $\Delta \lambda$ and $\Delta \phi$ are shown for these five databases in Fig. \ref{f4a}, and compared with the historical storms in the HURDAT2 database since 1920. 
In contrary to the frequency distributions shown in Fig. \ref{f4}, the PDFs are in much better agreement with the HURDAT2 storms. 
However, as in the frequency distributions, the range of $\Delta \lambda$ and $\Delta \phi$ for the simulated databases are shorter, and the PDFs are underpredicted in the extreme ends of the distributions obtained for the HURDAT2 storms. 

\begin{figure}[!ht]
    \begin{center}
        \includegraphics[trim=2cm 5cm 2cm 5cm,clip,width=0.65\textwidth]{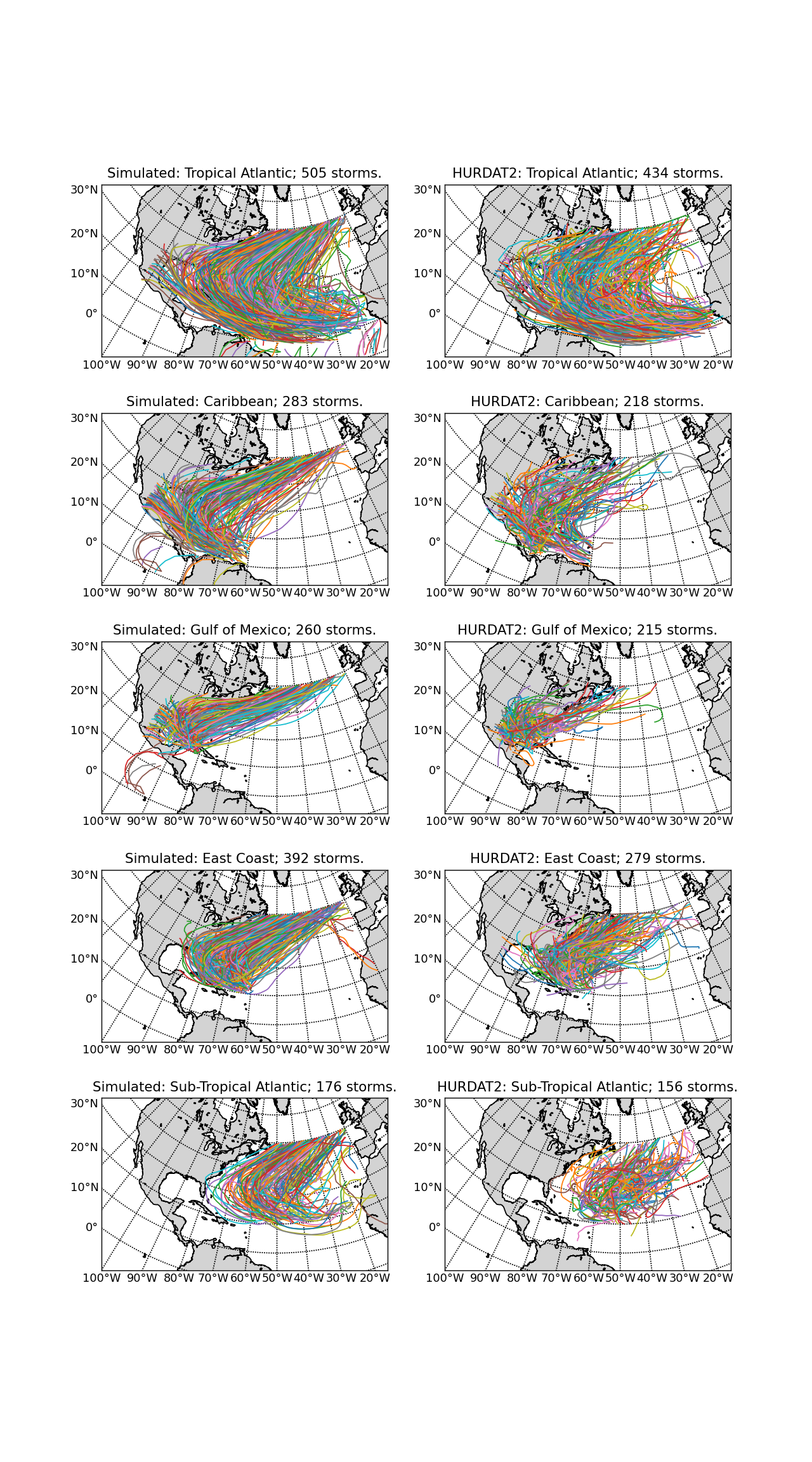}    
        \caption{Comparison of storm trajectories based on their inception sub-basin: simulated storms for an arbitrary $100-yr$ period (left); the HURDAT2 storms since 1920 (right).}
        \label{f7}
    \end{center}
\end{figure}

The trajectories of storms generated in each of the five sub-basins from one of the five simulated storm databases are compared with those of the HURDAT2 database in Fig. \ref{f7}. 
The frames in the left and right columns show the storm trajectories from the simulated database and from the HURDAT2, respectively. 
Excellent visual similarity is apparent between the trajectories of storms generated in all sub-basins; although, 
the trajectories of the simulated storms are smoother than the HURDAT2 storms even after adding the error terms in Eqs. \ref{lam1} -- \ref{phi2}. 
A few synthetic storms generated in the Caribbean and the Gulf of Mexico sub-basins and in the Pacific ocean travel westward, and some storms generated in the tropical Atlantic (close to the equator) travel southward. 
These anomalies are expected because the initial/genesis coordinates of these storms fall out of the range of the coordinates listed in HURDAT2, and therfore are not available for model training. 
Neural networks are known to provide erroneous results for out-of-bound inputs. The movements of these storms show up as the anomalous time records in Fig. \ref{f5} at the left end of $\phi$. 

Landfall status is an important derived property of the trajectory models that is pertinent to accurate estimation of windspeeds on the US mainland. 
In the HURDAT2 database, 370 storms made landfall in the US out of the 1302 storms listed since 1920 (i.e., $\sim 28.4\%$ of storms). 
For the five simulated storm databases presented herein, each for a $100-yr$ period with an average of 1619 storms per database, the average number of storms making landfall in the US was 480.2 (i.e., $\sim29.6\%$ of storms). 
The good agreement between proportion of simulated and actual storms that reach the mainland further validates the accuracy of the trajectory models and justifies that they are fit for the purpose estimating hurricane induced extreme coastal wind speeds. 

\subsection{Efficacy of the intensity models}

The trajectory and the intensity models are coupled via their input features; 
therefore, their efficacy cannot be assessed separately. 
In this section, we demonstrate that the predictions from the intensity models, i.e., $p_c$ and $w_m$, faithfully emulate the same characteristics in HURDAT2. 
To account for the larger number of storms in the simulated storm databases PDFs are preferred over frequency distributions for statistical comparison purposes. 

\begin{figure}[!ht]
    \begin{center}
        \includegraphics[trim=0cm 0cm 0.0cm 0cm,clip,width=0.75\textwidth]{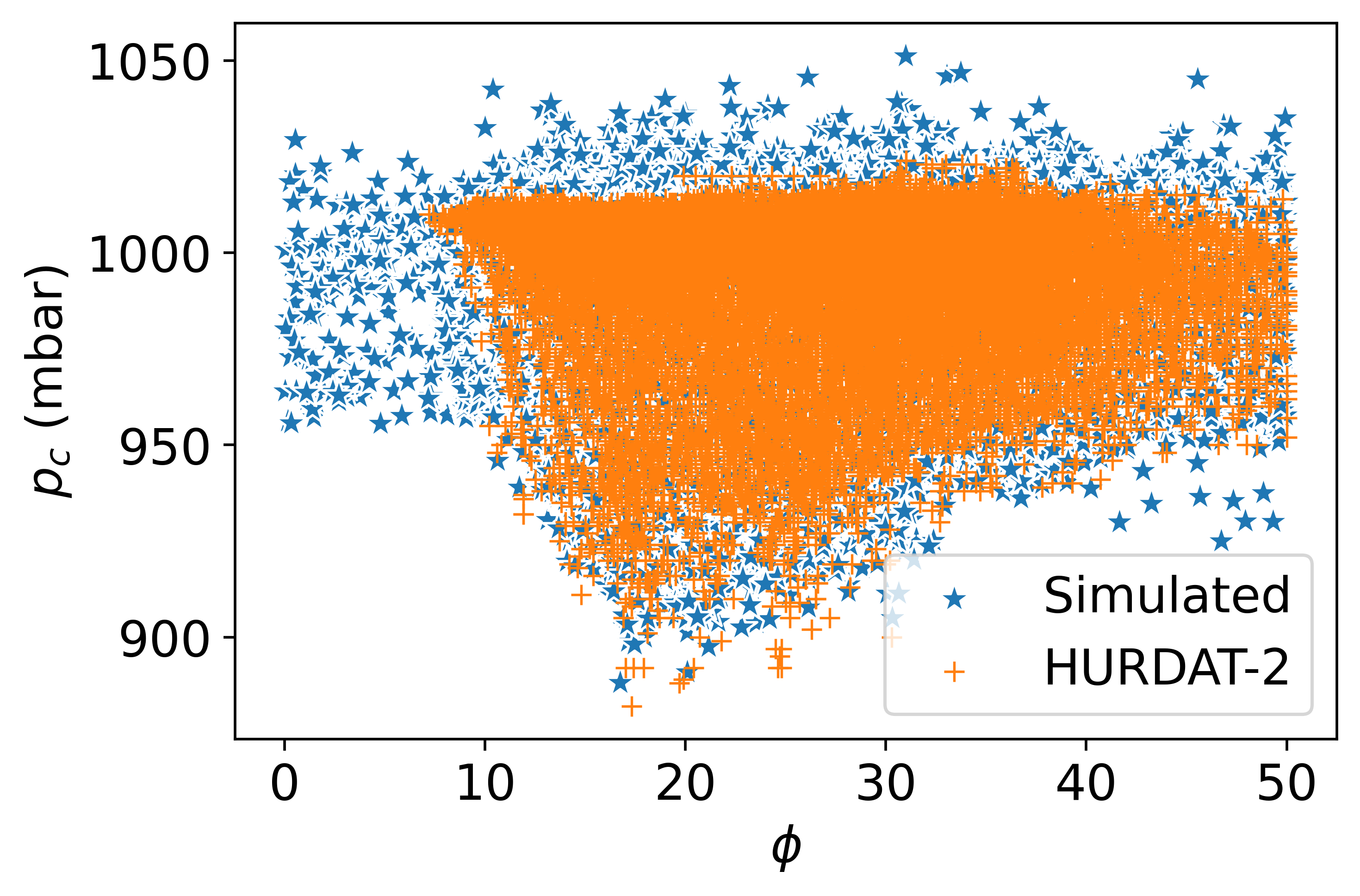}  
        \includegraphics[trim=0cm 0cm 0.0cm 0cm,clip,width=0.75\textwidth]{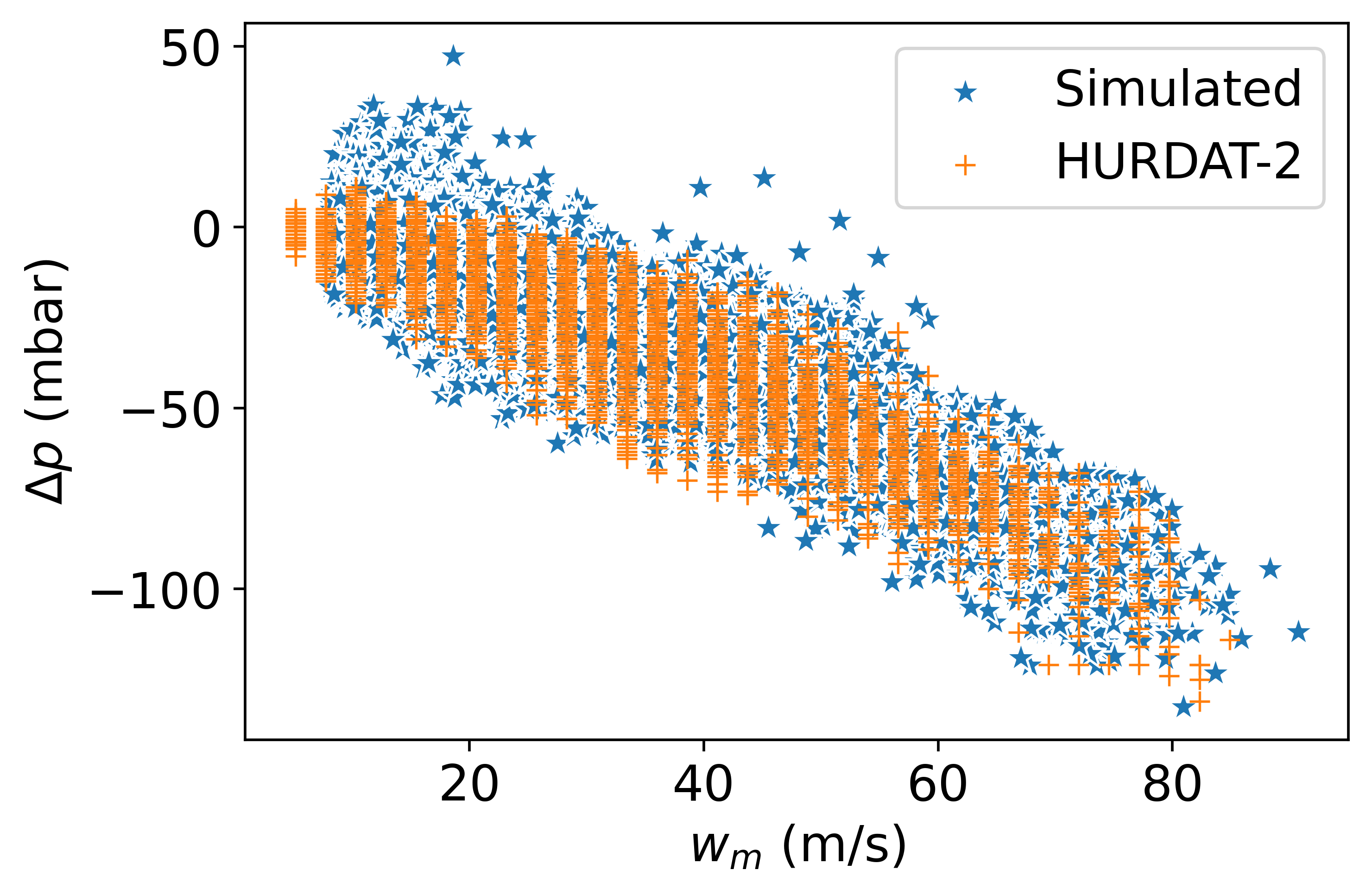}    
        \caption{Top frame: Distribution of central pressure ($p_c$) plotted against the latitude $\phi$ from a simulated storm database for a $100-yr$ period compared with that from the HURDAT2 database since 1920.
        Bottom frame: Joint scatterplots of pressure deficit ($\Delta p = p_c - p_n$) and the max. wind speed $w_m$ from a simulated storm database for a $100-yr$ period compared with that from the HURDAT2 database since 1920.} 
        \label{f8}
    \end{center}
\end{figure}

In the top frame of Fig. \ref{f8}, the central pressure ($p_c$) predicted by the intensity models is plotted against latitude ($\phi$) for one of the simulated storm databases and is compared with the HURDAT2 storms since 1920. 
The simulated $p_c(\phi)$ values are in good agreement with the HURDAT2 storms. 
The lowest values of $p_c$ in HURDAT2 are obtained in an intermediate range of $\phi \in [12^\circ N, 30^\circ N]$. 
The synthetic storms captures this property. 
Additionally, the synthetic storms accurately capture the spread in the $p_c$ at all $\phi$. 
The time records corresponding to the anomalous southward traveling storms are clearly discernible in top panel of Fig. \ref{f8} at low $\phi < 10^\circ N$, for which the intensity models assign values in the median range of $p_c$ from the HURDAT2 database.  

The pressure deficit $\Delta p = p_c - p_n$, where $p_n$ is the ambient pressure ($= 1013$ $mbar$) is plotted against $w_m$ for one of the five simulated storm databases, and compared to the storms in the HURDAT2 database since 1920 in the lower panel of Fig. \ref{f8}. 
These quantities are clearly negatively correlated, and the simulated storms emulate the characteristics from HURDAT2 well. 
The spread in $\Delta p$ at a given $w_m$, and the spread in $w_m$ at a given $\Delta p$, are accurately captured by the simulated storms. 
Moreover, the models are able to predict high values of $w_m$ commensurate with HURDAT2. 
Therefore, the sequential application of separate models for $p_c$ and $w_m$ accounts for the intensity of the simulated storms effectively. 

\begin{figure}[!ht]
    \begin{center}
        \includegraphics[trim=0cm 0cm 0.0cm 0cm,clip,width=0.75\textwidth]{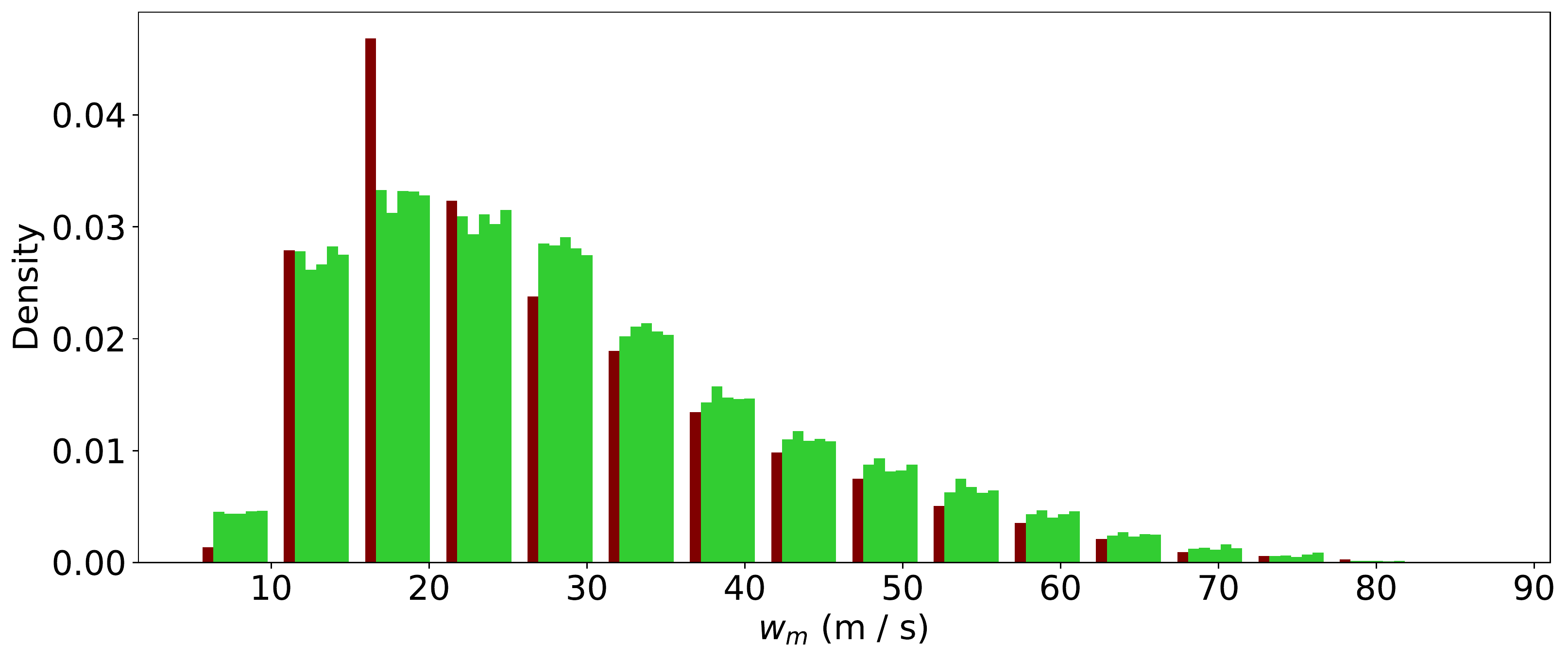}  
        \includegraphics[trim=0cm 0cm 0.0cm 0cm,clip,width=0.75\textwidth]{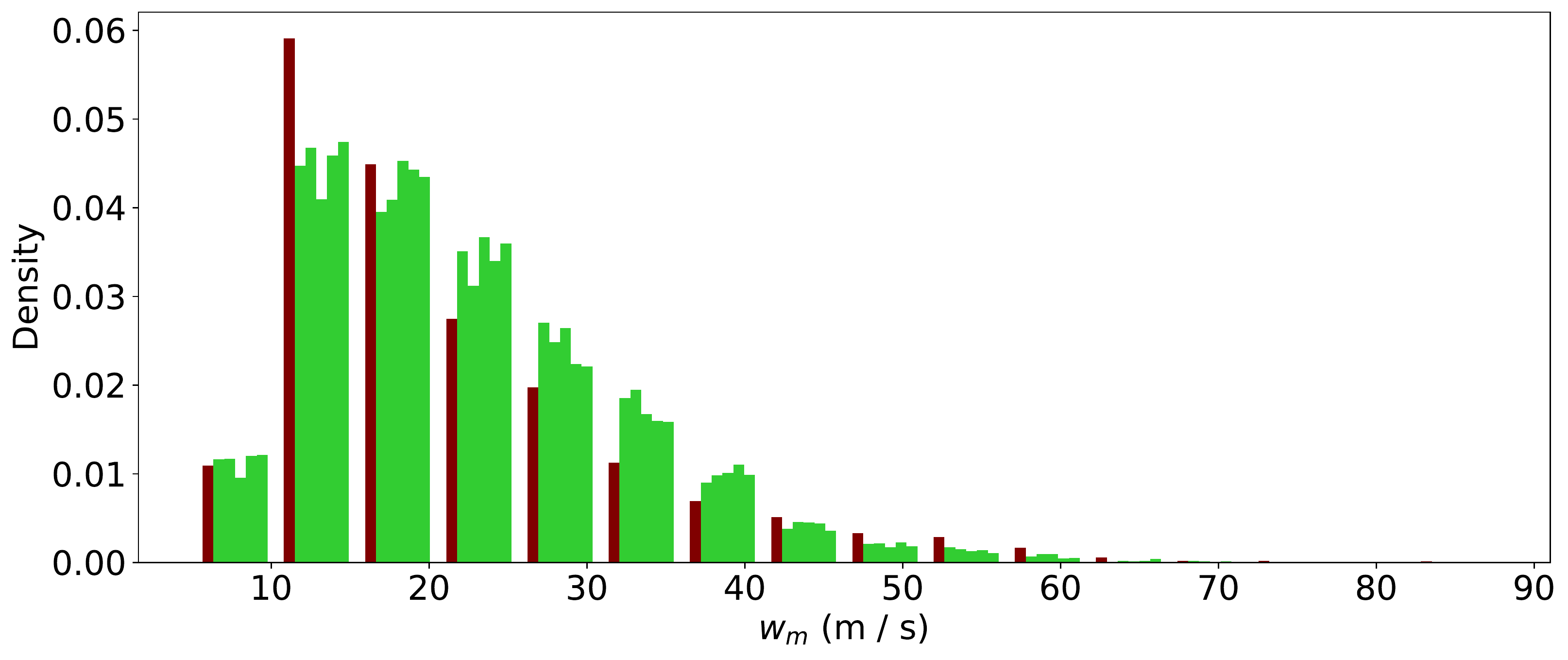}  
        \caption{Top frame: PDFs of the max. wind speed $w_m$ for five simulated storm databases, each simulated over a $100-yr$ period (green) compared with the same from the HURDAT2 database storms since 1920 (maroon).
        Bottom frame: PDFs of the max. wind speed $w_m$ for records with landfall status for five simulated storm databases, each simulated for a $100-yr$ period (green) compared with that from the HURDAT2 database storms since 1920 (maroon).}
        \label{f9}
    \end{center}
\end{figure}

The PDFs of $w_m$ for the five sets of synthetic storms, each for a period of $100$ $yrs$ and for the HURDAT2 storms since 1920 are compared in the top frame of Fig. \ref{f9}. 
The bottom frame of Fig. \ref{f9} compares the distributions of $w_m$ for the time records with landfall status in the US. 
The proportion of wind speeds greater 25 m/s is slightly overpredicted by the five sets of simulated storms.  The overprediction decreases as wind speed increases, and may be due to the use of the DB-3 for storm intensity modelling to emphasize the high wind speed storms in the training database, which are more pertinent for extreme wind estimation purposes. 
Both wind speed distributions (over land and overall) for the simulated storms are in very good agreement with HURDAT2 despite the discrepancies in the distributions of storm lifespan. 
The proportion of time steps with predicted $w_m \ge 75$ $ms^{-1}$ in the top frame and $w_m \ge 50$ $ms^{-1}$ in the bottom frame is slightly lower than observed in HURDAT2.
However, the under-prediction of the proportion of very high wind speeds over land is quite small, and is probably due to a combination of smoothing and the large number of low wind speed records over land in DB-3. 

\section{Local wind speed estimates}\label{sec4}

In this section, our simulation results are compared with HURDAT2 storms since 1920 in terms of the cumulative exceedance probability of the max.\ wind speed ($w_m$) recorded at three major locations in US mainland affected by destructive hurricanes. The trajectories of the storms making landfall near/ around those sites are also examined. 
The locations are Miami, Florida, New Orleans, Louisiana, and the Cape Hatteras in North Carolina.
Miami and New Orleans are chosen because they are severely affected by storm surge. 
Additionally, they are affected by storms with different trajectory trends. 
Storms originating in the tropical Atlantic and the Caribbean graze past/ over Miami, but storms from these sub-basins often make landfall in New Orleans.
Also, many storms making landfall at New Orleans originate in the Gulf of Mexico. 
Therefore, wind speed estimation at New Orleans is especially challenging. 
Cape Hatteras is located on the Hatteras Island, one of the barrier islands along the Atlantic shoreline, and is one of the most vulnerable regions for hurricanes.   

\begin{figure}[!ht]
    \begin{center}
        \includegraphics[trim=0cm 0cm 0.0cm 0cm,clip,width=0.75\textwidth]{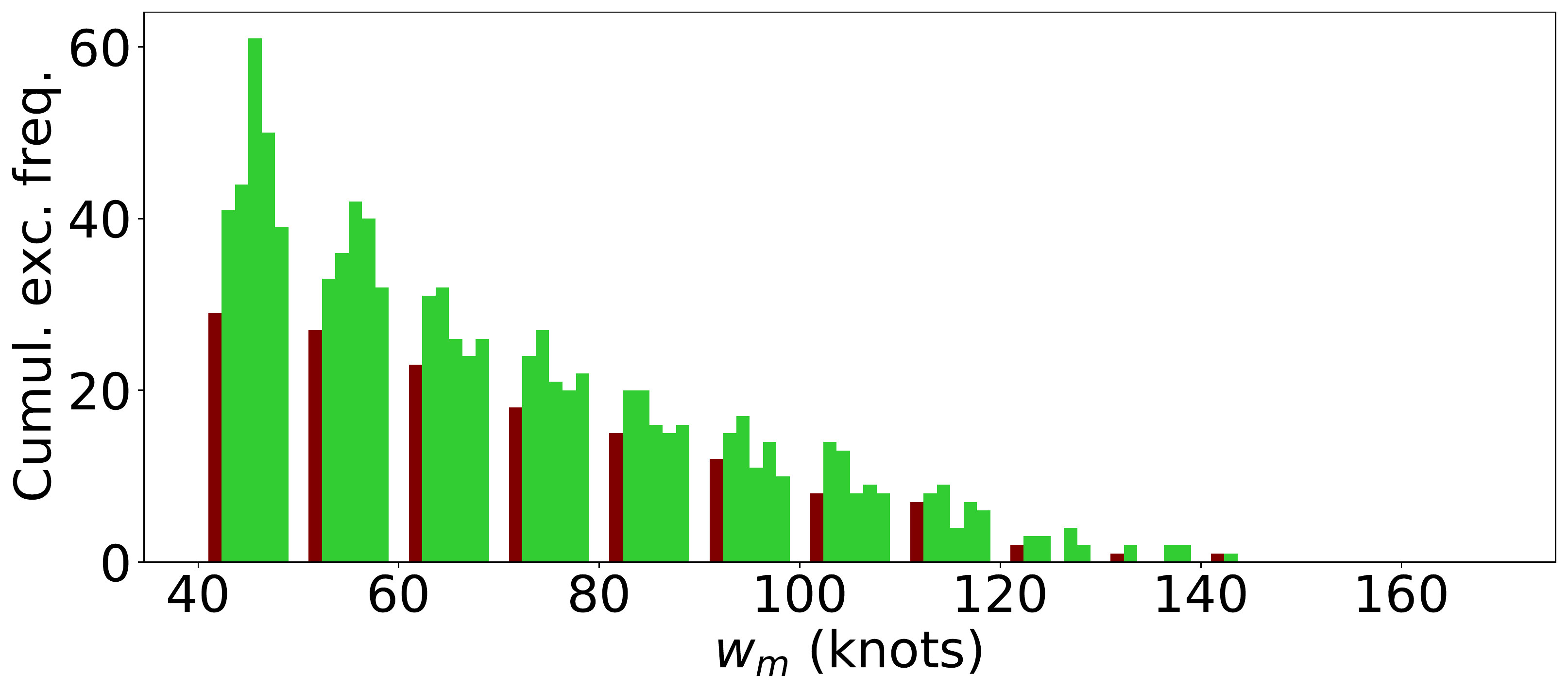}    
        \caption{Cumulative exceedance frequency distribution of the max. wind speed $w_m$ recorded for any storm passing within 100 $km$ of Miami computed for the simulated storm databases over five $100-yr$ periods (green) and compared with that from the HURDAT2 database since 1920 (maroon).}
        \label{f10}
    \end{center}
\end{figure}

\begin{figure}[!ht]
    \begin{center}
        \includegraphics[trim=0cm 0cm 0.0cm 0cm,clip,width=0.48\textwidth]{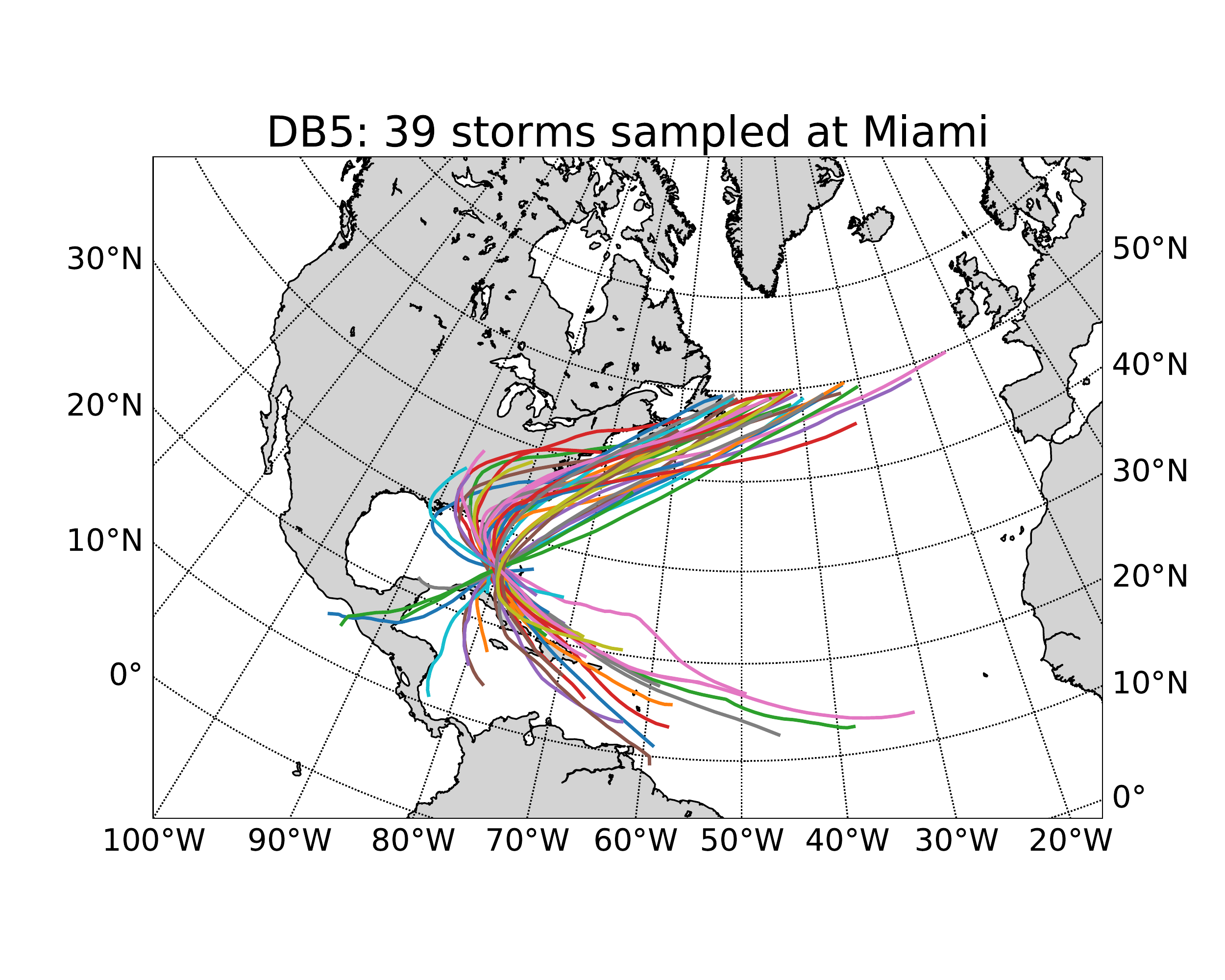}
        \includegraphics[trim=0cm 0cm 0.0cm 0cm,clip,width=0.48\textwidth]{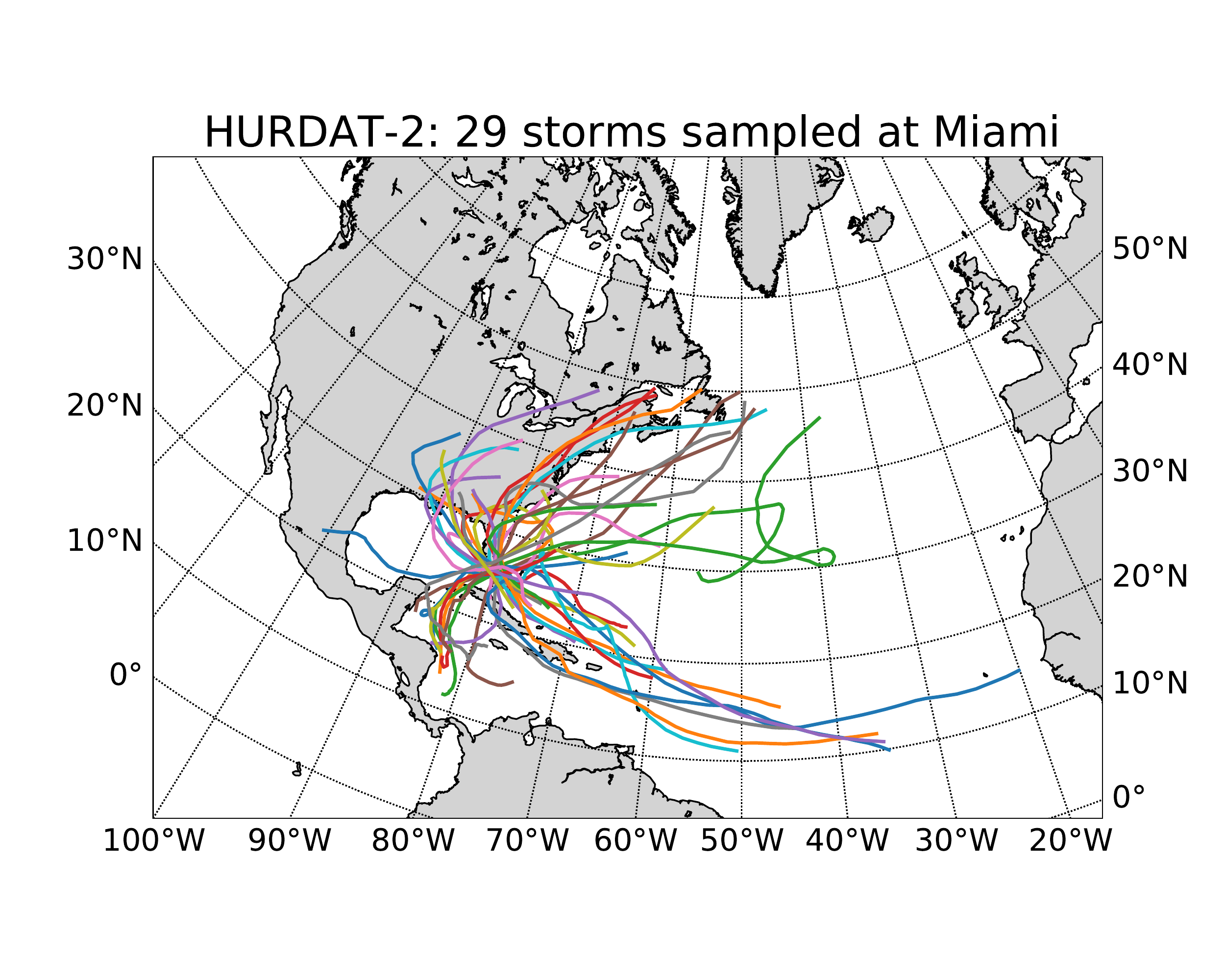}    
        \caption{Trajectories of storms passing within 100 $km$ of Miami for the simulated storms over an arbitrary $100-yr$ period (left) compared with those from the HURDAT2 database since 1920 (right).}
        \label{f11}
    \end{center}
\end{figure}

\begin{figure}[!ht]
    \begin{center}
        \includegraphics[trim=0cm 0cm 0.0cm 0cm,clip,width=0.4\textwidth]{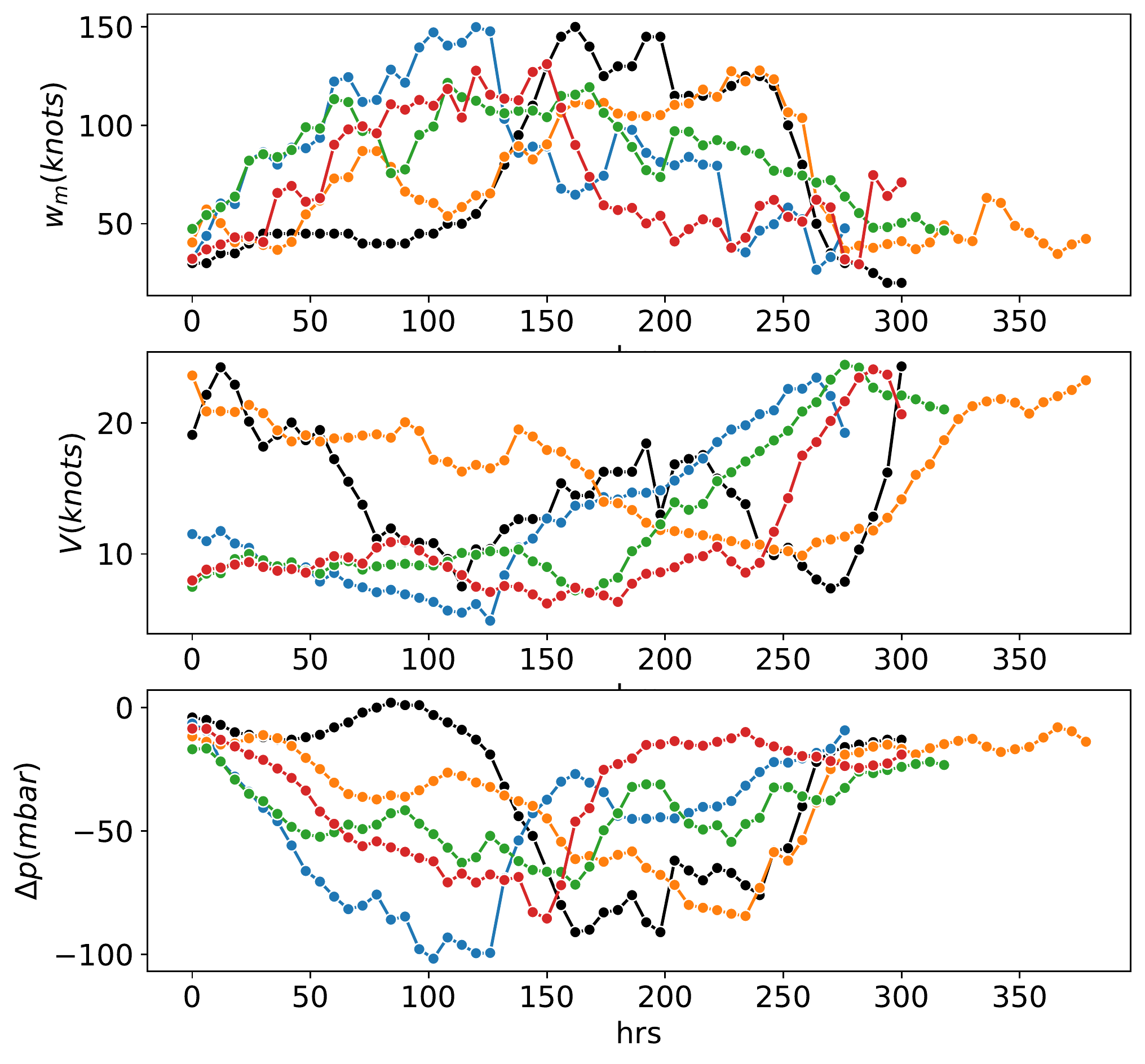}
        \includegraphics[trim=0cm 0cm 0.0cm 0cm,clip,width=0.48\textwidth]{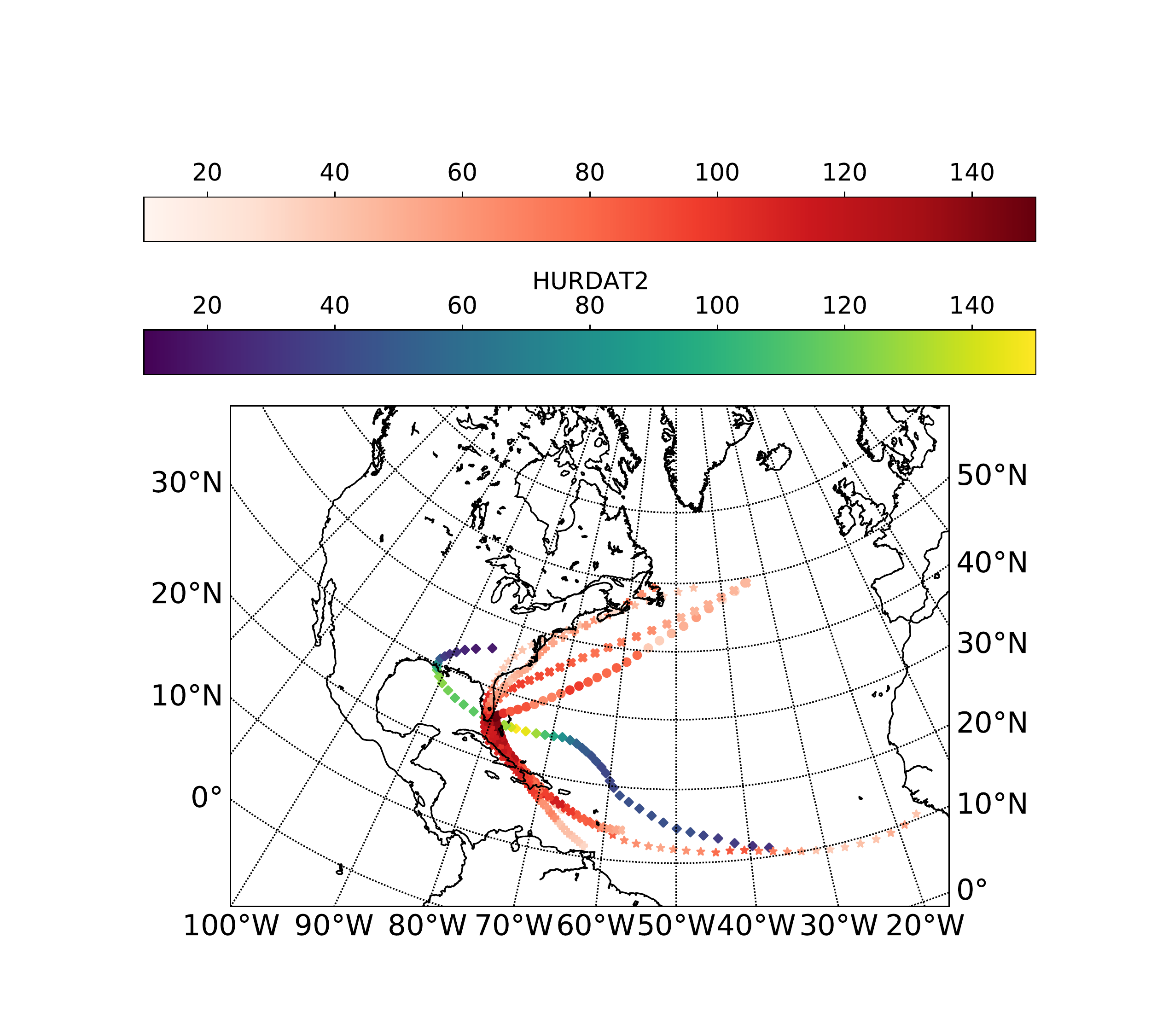}    
        \caption{Left: Max. wind speed $w_m$, translation speed $V$, and pressure deficit $\Delta p$ plotted over lifespan of the storms that induce maximum $w_m$ at Miami in each of the five simulated databases and the HURDAT2 storms (black symbols). Right: Trajectories of storms plotted in the left frame. 
        The colors in the trajectories indicate the wind speeds $w_m (knots)$.}
        \label{f11a}
    \end{center}
\end{figure}

\begin{figure}[!ht]
    \begin{center}
        \includegraphics[trim=0cm 0cm 0.0cm 0cm,clip,width=0.75\textwidth]{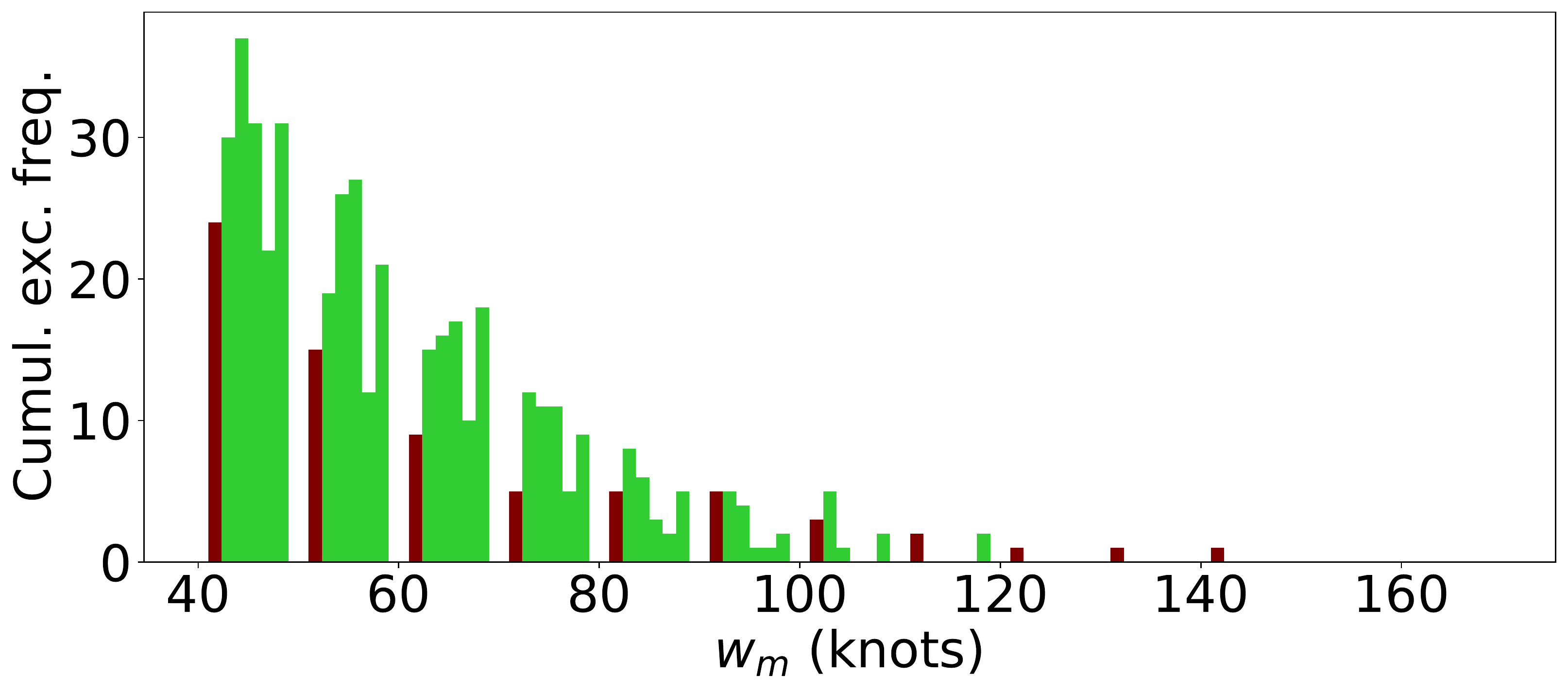}    
        \caption{Cumulative exceedance frequency distribution of the max. wind speed $w_m$ recorded for any storm passing within 100 $km$ of New Orleans computed for the simulated storms over five $100-yr$ periods (green) and compared with that from the HURDAT2 database since 1920 (maroon).}
        \label{f12}
    \end{center}
\end{figure}

\begin{figure}[!ht]
    \begin{center}
        \includegraphics[trim=0cm 0cm 0.0cm 0cm,clip,width=0.48\textwidth]{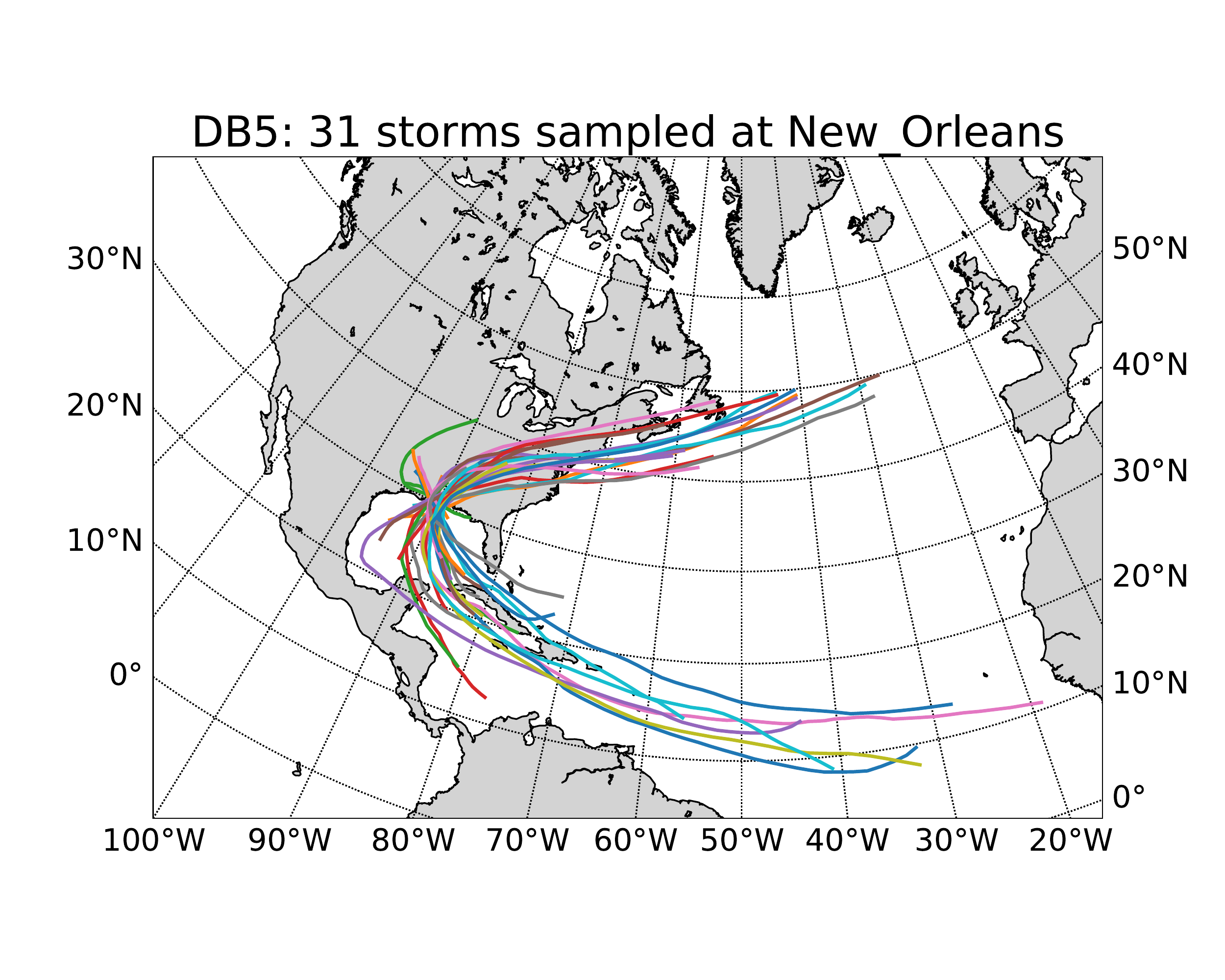}
        \includegraphics[trim=0cm 0cm 0.0cm 0cm,clip,width=0.48\textwidth]{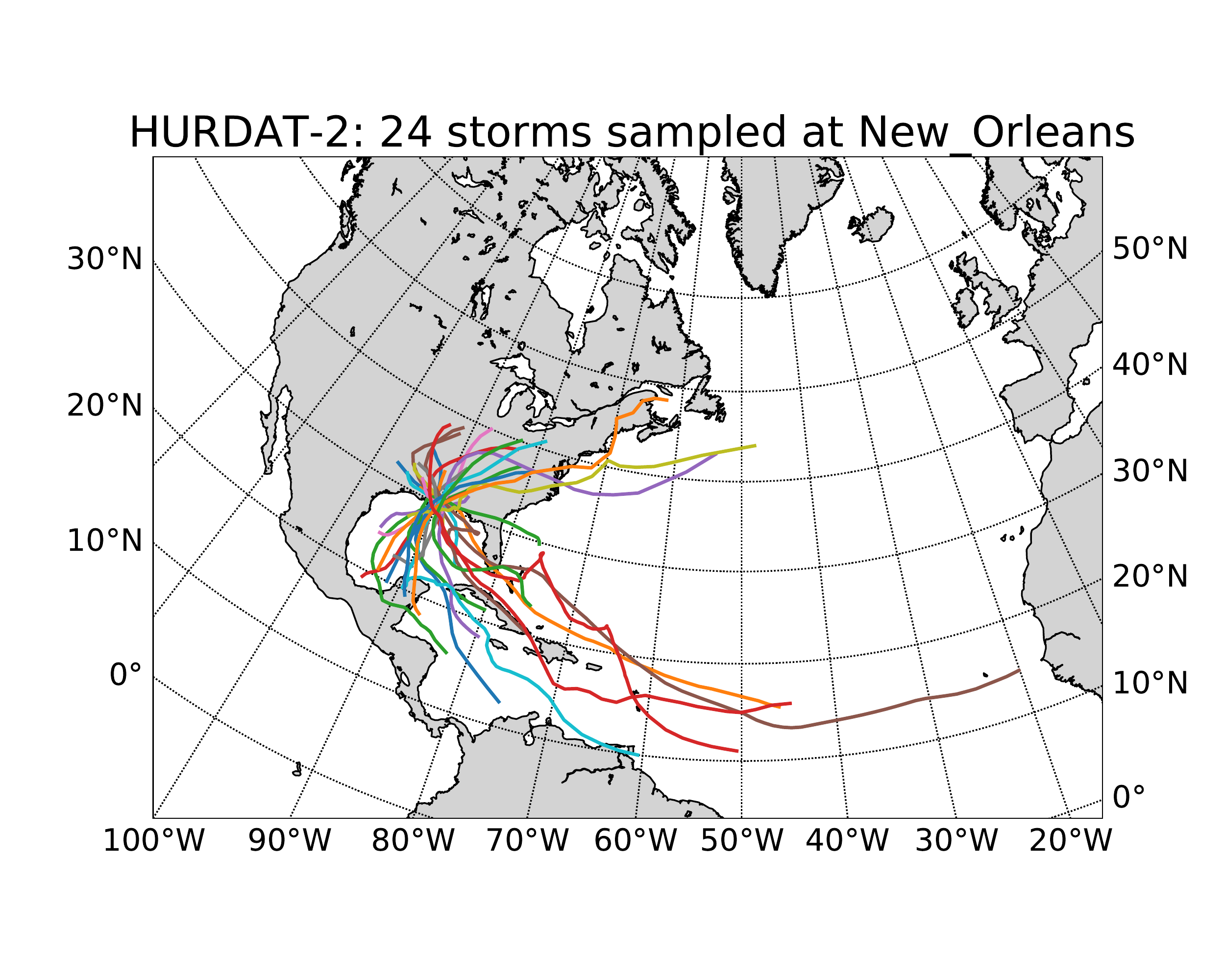}    
        \caption{Trajectories of storms passing within 100 $km$ of New Orleans for the simulated storms over an arbitrary $100-yr$ period (left) compared with those from the HURDAT2 database since 1920 (right).}
        \label{f13}
    \end{center}
\end{figure}

\begin{figure}[!ht]
    \begin{center}
        \includegraphics[trim=0cm 0cm 0.0cm 0cm,clip,width=0.4\textwidth]{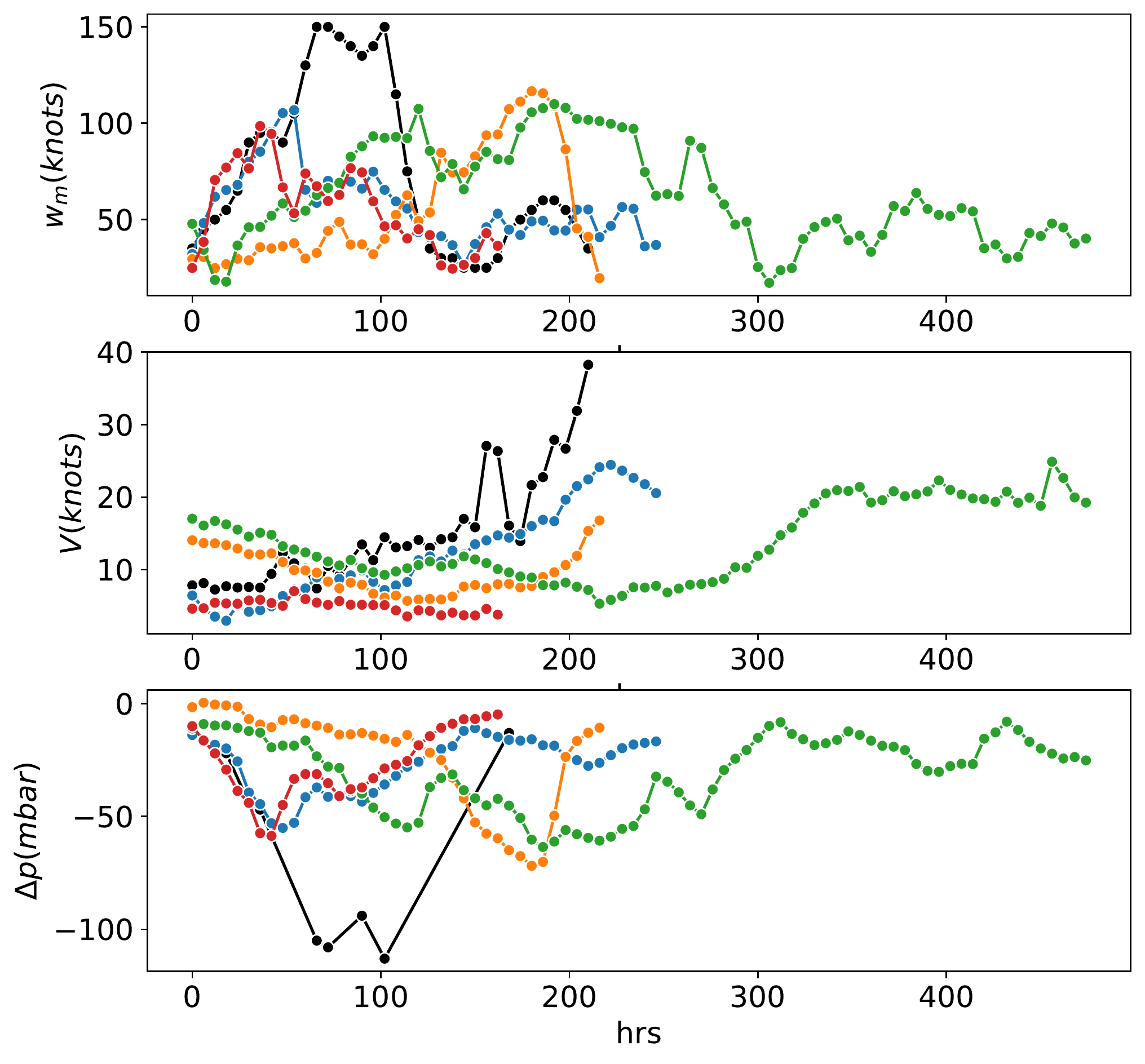}
        \includegraphics[trim=0cm 0cm 0.0cm 0cm,clip,width=0.48\textwidth]{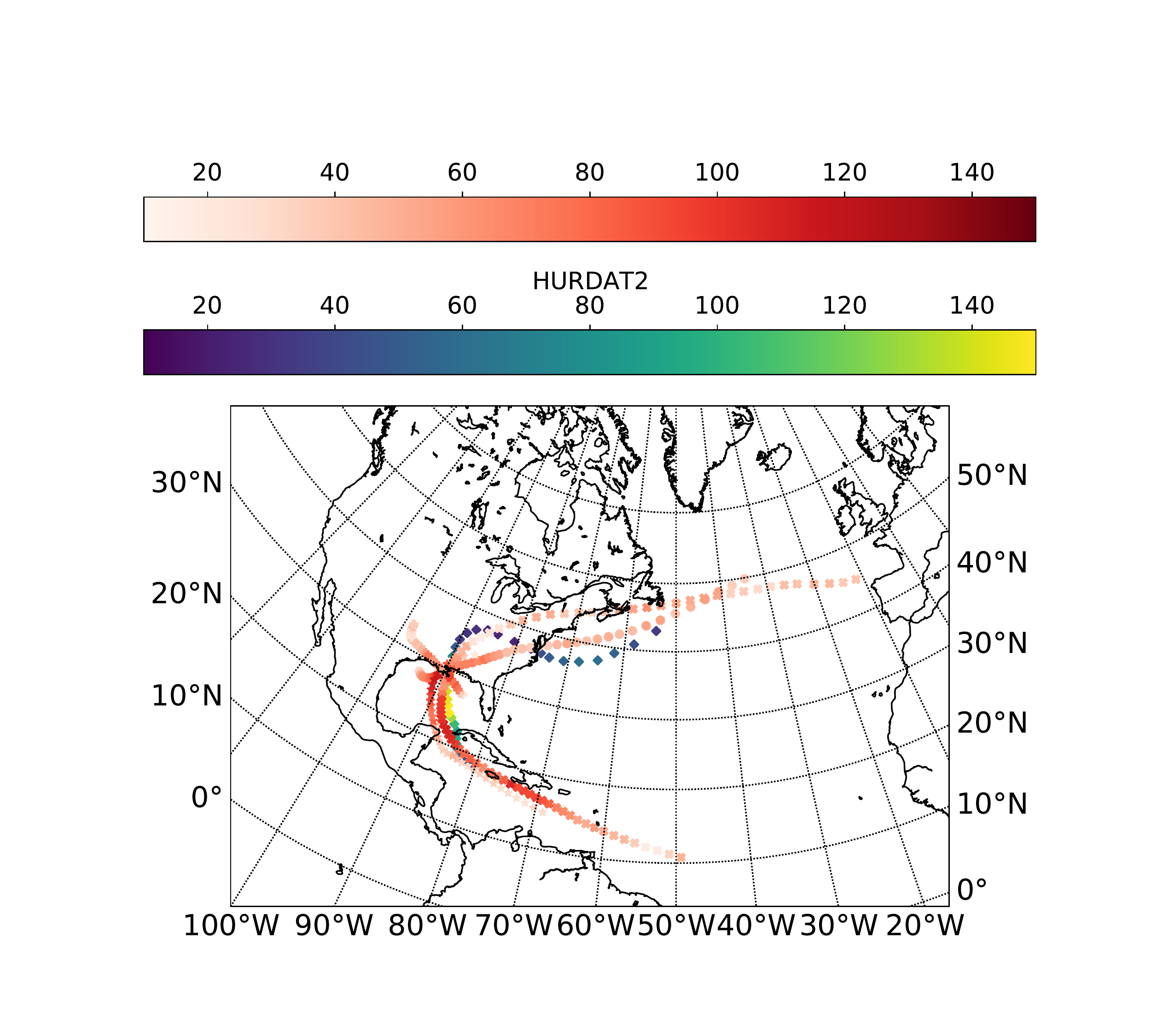}
        \caption{Left: Max. wind speed $w_m$, translation speed $V$, and pressure deficit $\Delta p$ plotted over lifespan of the storms that induce maximum $w_m$ at New Orleans in each of the five simulated databases and the HURDAT2 storms (black symbols). Right: Trajectories of storms plotted in the left frame. 
        The colors in the trajectories indicate the wind speeds $w_m (knots)$.}
        \label{f13a}
    \end{center}
\end{figure}

\begin{figure}[!ht]
    \begin{center}
        \includegraphics[trim=0cm 0cm 0.0cm 0cm,clip,width=0.75\textwidth]{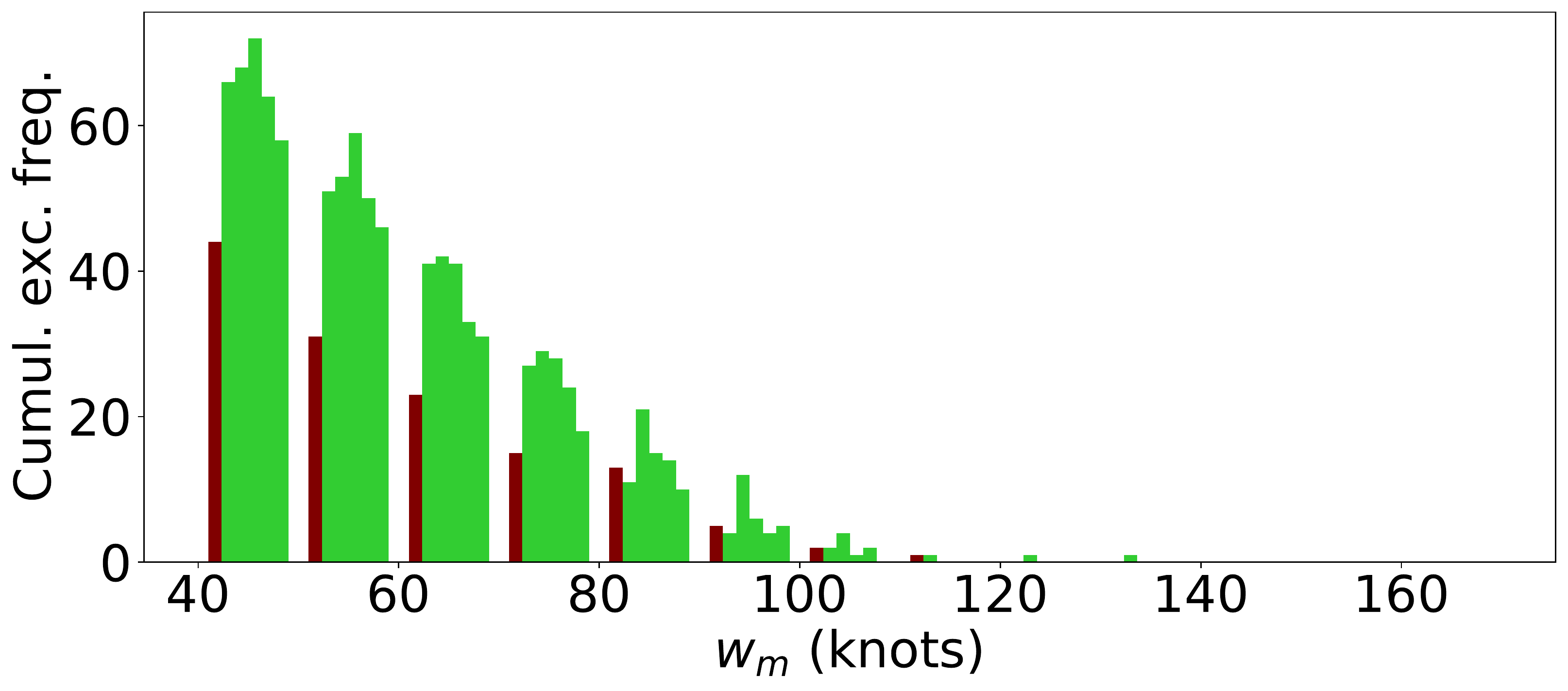}    
        \caption{Cumulative exceedance probability distribution of the max. wind speed $w_m$ recorded for any storm passing within 100 $km$ of Cape Hatteras computed for the simulated storms over five $100-yr$ periods (green) and compared with that from the HURDAT2 database since 1920 (maroon).}
        \label{f14}
    \end{center}
\end{figure}

\begin{figure}[!ht]
    \begin{center}
        \includegraphics[trim=0cm 0cm 0.0cm 0cm,clip,width=0.48\textwidth]{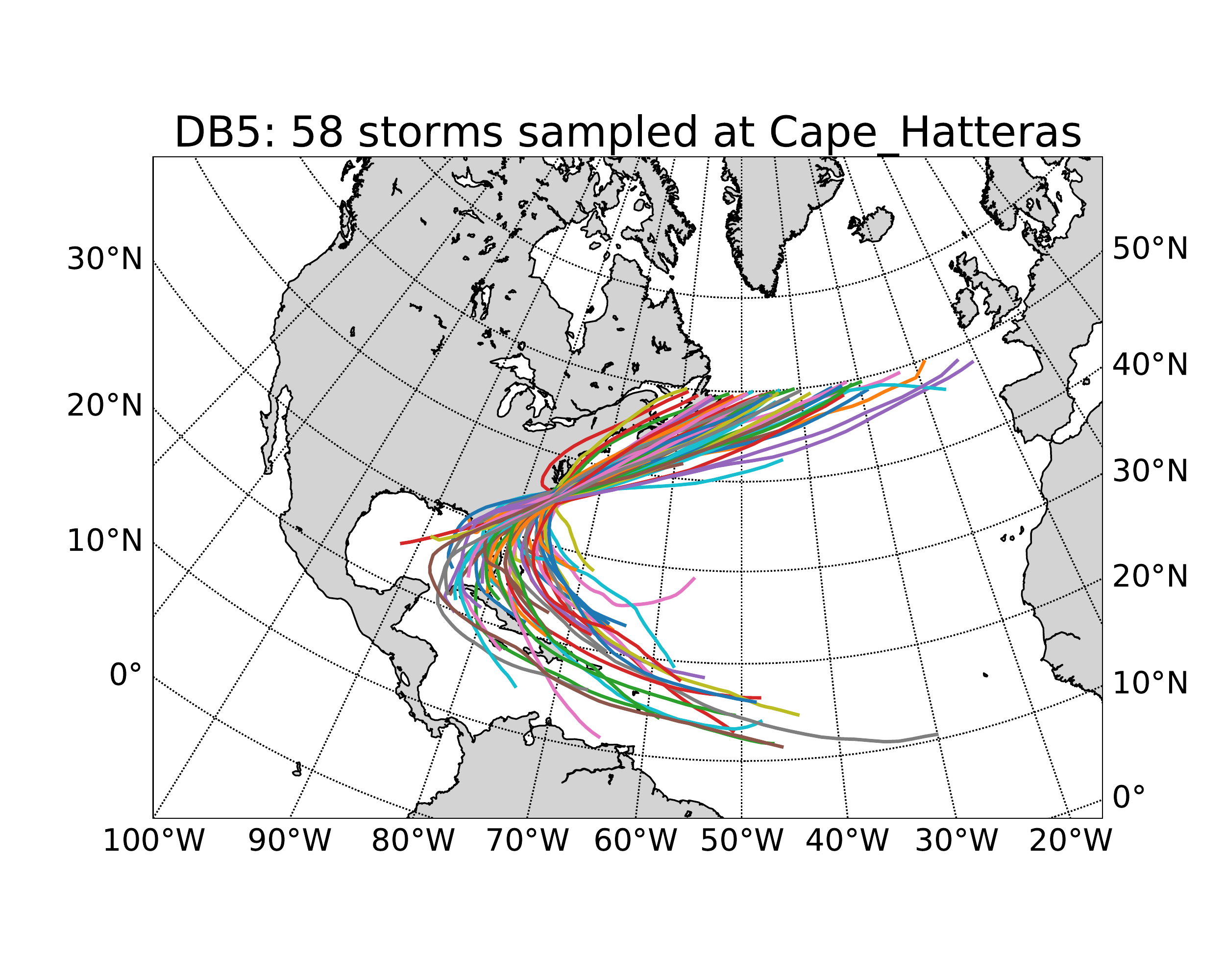}
        \includegraphics[trim=0cm 0cm 0.0cm 0cm,clip,width=0.48\textwidth]{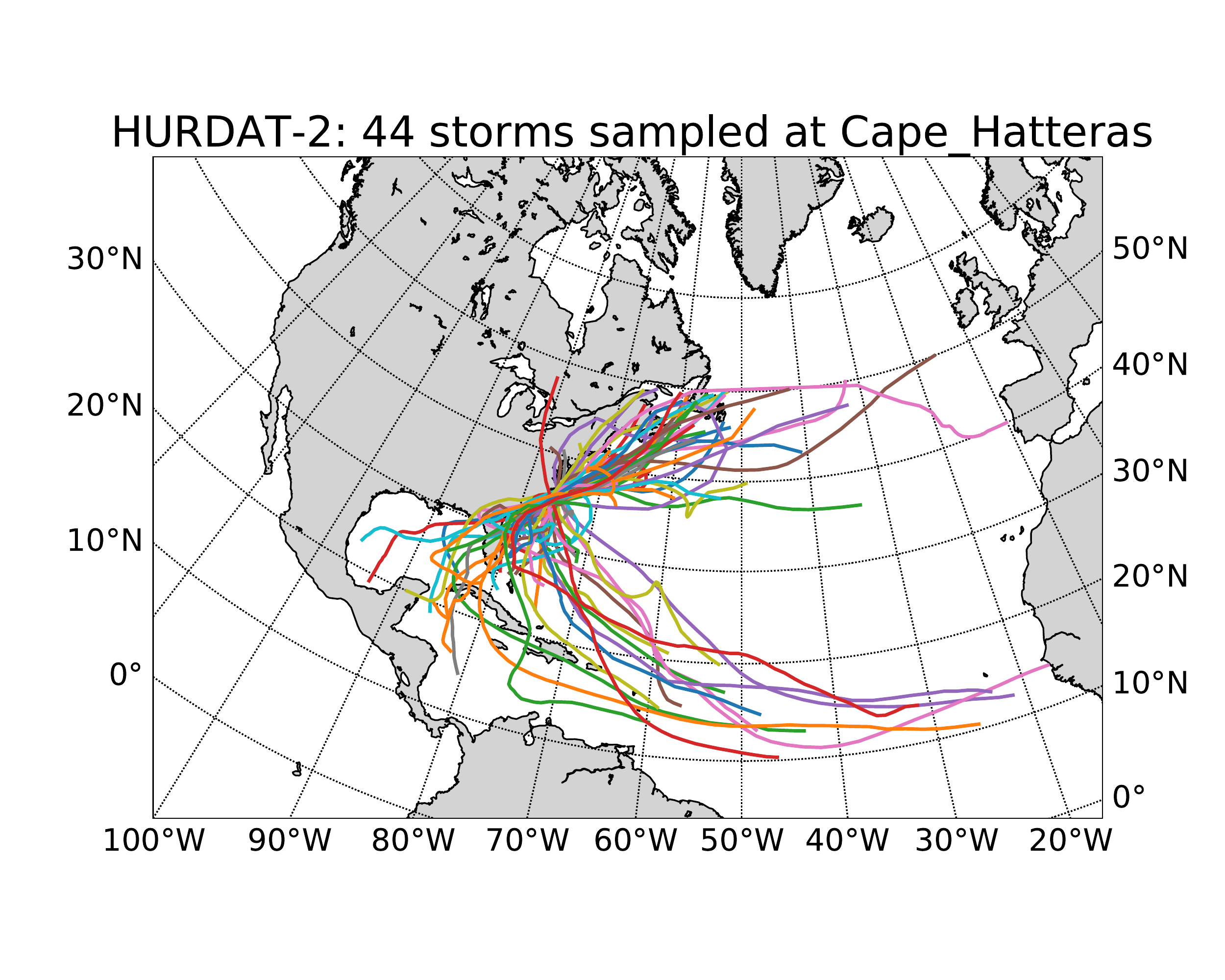}    
        \caption{Trajectories of storms passing within 100 $km$ of Cape Hatteras for the simulated storms over an arbitrary $100-yr$ period (left) compared with those from the HURDAT2 database since 1920 (right).}
        \label{f15}
    \end{center}
\end{figure}

\begin{figure}[!ht]
    \begin{center}
        \includegraphics[trim=0cm 0cm 0.0cm 0cm,clip,width=0.4\textwidth]{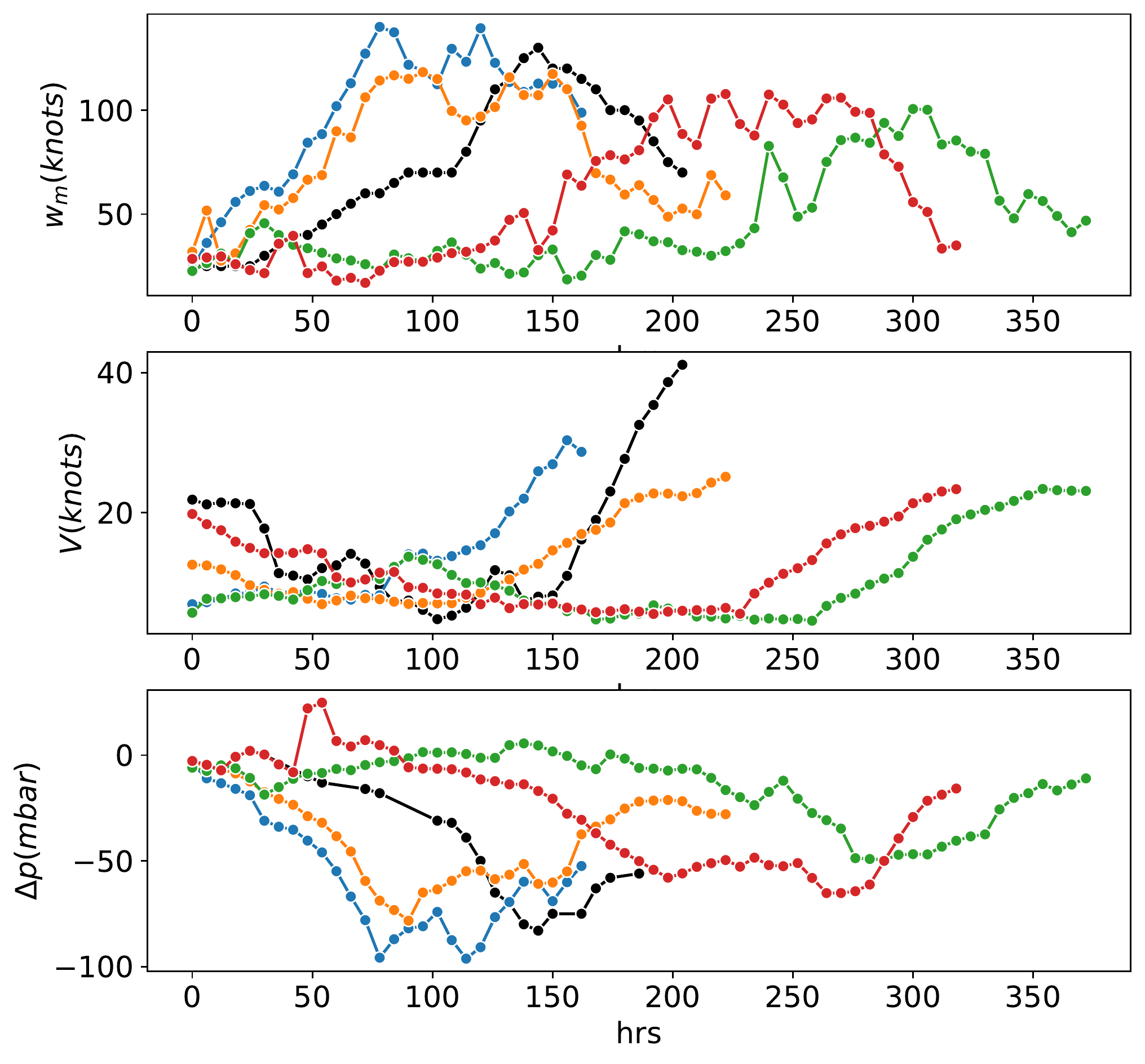}
        \includegraphics[trim=0cm 0cm 0.0cm 0cm,clip,width=0.48\textwidth]{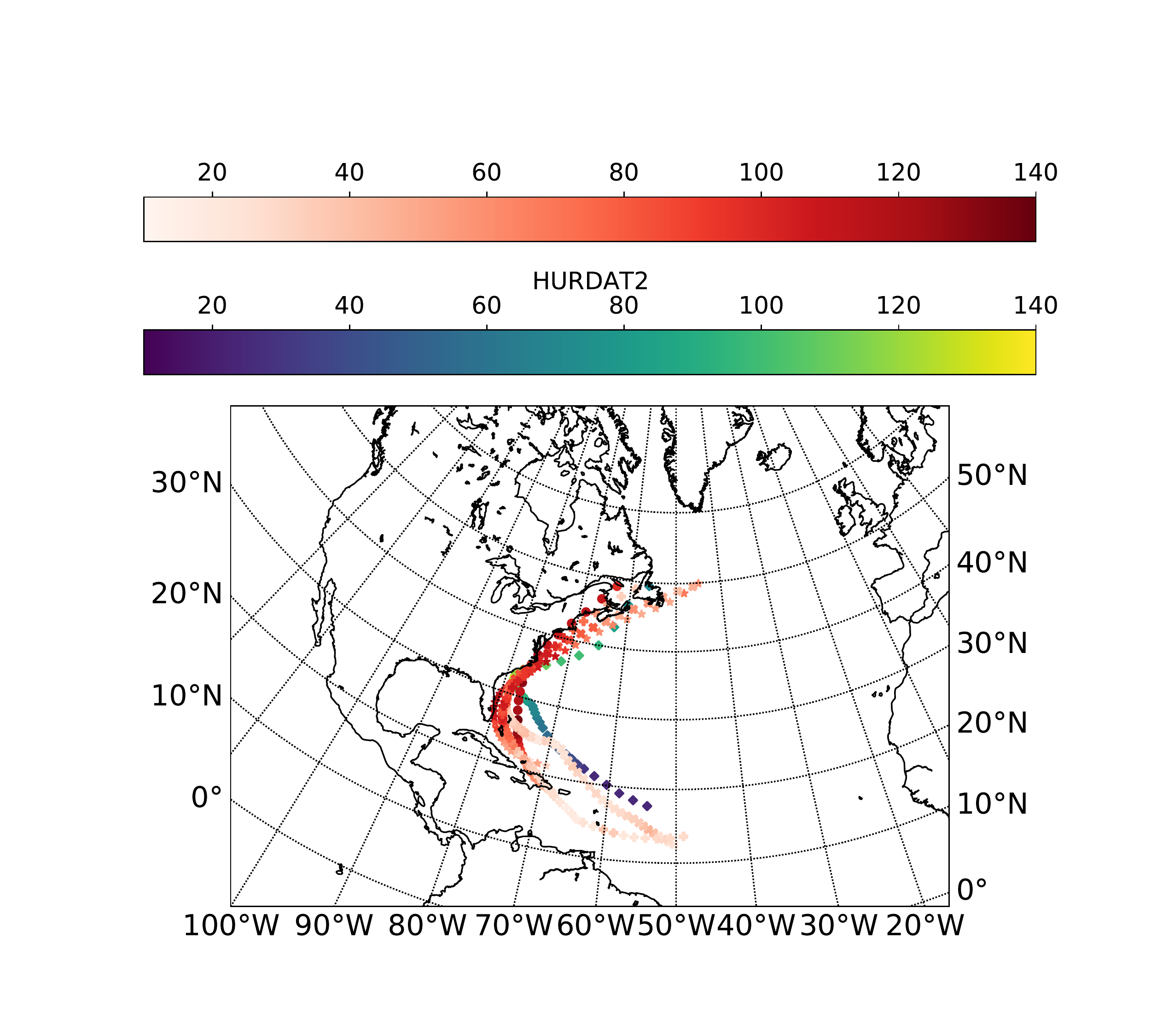}
        \caption{Left: Max. wind speed $w_m$, translation speed $V$, and pressure deficit $\Delta p$ plotted over lifespan of the storms that induce maximum $w_m$ at Cape Hatteras in each of the five simulated databases and the HURDAT2 storms (black symbols). Right: Trajectories of storms plotted in the left frame. 
        The colors in the trajectories indicate the wind speeds $w_m (knots)$.}
        \label{f15a}
    \end{center}
\end{figure}

Only storms that pass within 100 $km$ of a chosen site's location in spherical coordinates attaining $w_m\ge40$ $knots$ are considered. 
In Figs. \ref{f11}, \ref{f13}, and \ref{f15}, trajectories of storms satisfying those criteria are plotted for one of the five simulated databases (left frame) and compared with the trajectories of storms from HURDAT2 since 1920 (right frame). 
Figures~\ref{f10}, \ref{f12} and \ref{f14} show the cumulative exceedance probability of $w_m$ based on simulated storms and the storms in HURDAT2 since 1920 in $knots$ at Miami, New Orleans and Cape Hatteras, respectively. 
In Figs.~\ref{f10}, \ref{f12}, and \ref{f14}, the records from the HURDAT2 database are shown in maroon bars, and the frequencies from the five databases are all shown in green bars. 
In Figs. \ref{f11a}, \ref{f13a}, and \ref{f15a} the storm that induces the maximum $w_m$ at the chosen locations in each database is analyzed by plotting the storm intensity characteristics, $w_m$, $V$, and $\Delta p$ over the storm's lifespan in the left frames and their trajectories in the right frames. The five simulated storms are shown in color and the storm from HURDAT2 is shown in black in the characteristic plots, and 
in the storm trajectory plots, the symbols are colored by the value of $w_m$. 

There are 29 storms in HURDAT2 since 1920 which at some point in their lifetime pass within 100 $km$ of downtown Miami with $w_m \ge 40$ $knots$, as shown in the right frame of Fig. \ref{f11}. 
Considering the small number of time records available for comparison, the cumulative exceedance frequency of $w_m$ for the simulated storms plotted in Fig. \ref{f10} are in good agreement with the cumulative exceedance frequency of $w_m$ for the HURDAT2 storms. 
The exceedance frequencies for the simulated storms are over-predicted at low values of $w_m$ because of the larger number of simulated storms in each of the five synthetic storm databases. 
However, in the medium to high $w_m$ range, the exceedance frequencies for the five simulated storm databases are very similar to the HURDAT2 storms. 
Also, the trajectories of the simulated storms are in good agreement with the HURDAT2 storms in Fig. \ref{f11}. 
Storm headings on arrival to and departure from Miami are in good agreement. 
The storm trajectories of the simulated storms are smoother than the HURDAT2 storm trajectories due to smoothing associated with the use of the neural network trajectory models. 
Importantly, the high wind speeds induced by the storms generated in the tropical Atlantic and the Caribbean sub-basins in HURDAT2 are replicated in the simulated storm databases at approximately the correct frequencies. 

Storm intensity characteristics of the most powerful storm recorded within 100 $km$ of Miami in each of the five simulated databases and from HURDAT2 since 1920 are compared in Fig. \ref{f11a}. 
The six-hourly evolutions of $w_m$, $V$ and $\Delta p$ for the five simulated storms (colored symbols) are similar to evolution of the historical storm (black symbols) in the left frame of Fig. \ref{f11a}. 
Although the trajectory and intensity models emulate the historical storm evolution properties, the individual storms are distinctly different from each other. 
The storm lifespans also vary significantly. 
The highest $w_m$ induced at Miami by the simulated storms is 150 $knots$ compared to 145 $knots$ in HURDAT2 since 1920. 
High $w_m$ is generally induced by storms with $V < 25$ $knots$. 
The correlation between $w_m$ and $\Delta p$ for the simulated storms is similar to the historical storm. 
The trajectories of the storms colored by $w_m$ in the right frame of Fig. \ref{f11a} show that the historical storm trajectory trends are indeed replicated by the DL trajectory models. 
Interestingly, as is the case for the HURDAT2 storm, the most destructive simulated storms recorded at Miami are generated in the tropical Atlantic sub-basin. 
However, the trajectory of the HURDAT2 storm is qualitatively different from the simulated storms. 
Spatial distribution of high wind speeds from the simulated storms is in good agreement with the historical storm. 

The cumulative exceedance frequencies of $w_m$ for the storms passing within 100 $km$ of New Orleans are shown in Fig. \ref{f12}. At New Orleans, the exceedance frequencies of $w_m$ for the simulated storms are more similar to those from HURDAT2 in the low $w_m$ range as compared to Miami. 
On the other hand, in the high windspeed range, $w_m \ge 95$ $knots$, the exceedance frequencies of $w_m$ for the simulated storms tends to be smaller than from the HURDAT2 storms.
The prediction of the very high wind speeds from HURDAT2 is more complicated in the Gulf region. 
This is the case because storms heading northward in this region make landfall and quickly dissipate, causing a steep drop off in $w_m$ for storm trajectories in the training dataset. 
The smoothing inherent in RF models predicts a more gradual drop-off in $w_m$, and therefore, typically lower wind speeds are predicted by the RF intensity models along the Gulf coastline even before landfall. 
There are fewer storm records with $w_m \ge 40$ $knots$ for New Orleans as compared to Miami, shown in Fig. \ref{f13}.  Only 24 storms in HURDAT2 since 1920 satisfy this criterion. 
However, as was observed at Miami, the trajectory trends are in remarkably good agreement with the HURDAT2 storm trajectories, even for the storms originating in the Gulf of Mexico. 
Additionally, the simulated storm trajectories post landfall are in good agreement with the HURDAT2 storms traversing over land. 

The evolution of storm intensity properties, at six-hour intervals, for the most destructive storm recorded within 100 $km$ of New Orleans from each of the five simulated databases are compared with the historically strongest storm recorded in New Orleans since 1920 in the left frame of Fig. \ref{f13a}. 
The lifespan of the historical storm is shorter than the simulated storms. 
However, the highest $w_m$ recorded at New Orleans by the historical storm ($w_m = 149.98$ $knots$) is larger than the highest $w_m$ recorded by any of the five simulated storms. 
The highest wind speed, $116$ $knots$, induced by the simulated storms is close to the second highest wind speed recorded in HURDAT2 since 1920, 115 $knots$. 
Again, the highest value of $w_m$ is typically induced when the storm translation speed, $V$, is low ($< 15$ $knots$). 
As expected, $w_m$ and $\Delta p$ are found to be negatively correlated to each other. 

The storm trajectories shown in the right frame of Fig. \ref{f13a} show that, unlike Miami, the most destructive simulated storms at New Orleans arise from any of the tropical Atlantic, Caribbean, or the Gulf of Mexico sub-basins. 
The spatial distribution of high-$w_m$ records is also captured by the simultated storms.
The inherent dissipative characteristics of the intensity models over land are visually evident in the right frame of Fig. \ref{f13a} by the reduction in $w_m$ as the simulated storms make landfall. 

The cumulative exceedance frequencies of $w_m$ at Cape Hatteras are representative of the efficacy of the simulation model in generating storms that graze past the east coast of the US mainland. 
Good agreement between cumulative exceedance frequencies of $w_m$ for the simulated storms and the HURDAT2 storms is evident in Fig. \ref{f14}. 
Again the simulated storm databases show higher cumulative exceedance frequencies of $w_m$ than the HURDAT2 storms since 1920 in the low to medium ranges of $w_m$ owing to the larger number of storms in the simulated databases (also see Fig. \ref{f15}). 
However, the models successfully predict high wind speeds at approximately the right frequencies, which is the most important consideration for purpose of this work. 
The storm trajectories of the simulated storms affecting Cape Hatteras are also in good agreement with the HURDAT2 storms (Fig. \ref{f15}); although, the simulated storm trajectories are smoother. 

The evolution of intensity properties at six-hour time steps for the most powerful simulated storms (colored symbols) at Cape Hatteras and the most destructive historical storm since 1920 (black symbols) are shown in Fig. \ref{f15a}. 
Significant variation in lifespan of the simulated storms is noticeable. 
The highest $w_m$ recorded at Cape Hatteras for the simulated storms is in good agreement with the historically highest wind speed. 
Again, high $w_m$ is occurs when $V$ is low, and the simulation models incorporate adequate variability. 
The storm trajectories in the right frame of Fig. \ref{f15a} show that the pertinent trajectory trends are emulated by the trajectory models and the intensity models are able to provide the correct spatial distribution of the highest wind speeds.

The five simulated databases reflect the repeatability of and variability between the simulations. 
The results indicate that the simulations are repeatable and the variability between simulations is realistic by comparison to historical data. 
In our experience of simulating synthetic storms using models trained on HURDAT2 storms, 
estimation of extreme wind speeds over the east coast is easier compared to the southern Gulf coastline. 
This is due to a storm's bearing being typically acute to the coastline along the east coast at landfall and the bearing being near-normal along the southern Gulf coastline.



\section{Discussion \& Conclusions}\label{sec5}

This paper demonstrates the applicability of DL and ML techniques for estimating hurricane wind speeds. 
Recent success of ML/ DL methods in several aspects of atmospheric science and weather forecasting \citep{reichstein2019deep, schultz2021can} provided the impetus for this research. 
Storm trajectory models developed in \citet{bose2021forecasting} and \citet{bose2022real} using LSTM-RNN model architectures and $6-hr$ storm displacement probabilities as input features are also leveraged here. 
These models incur errors similar to those inherent in ensemble models used by the NHC for trajectory forecasts up to 12 hours. 
A set of intensity models were developed using RFs to complete the synthetic storm generation. 
These intensity models predict a storm's central pressure, $p_c$, and maximum wind speed, $w_m$. 
Previous works typically modelled storm intensities zonally, as local ambient conditions affect the evolution of a storm's intensity. 
In this work, alternatively, three sets of intensity models based on storm intensity ($w_m$) were used.
These intensity models have the advantage of embedding storm dissipation characteristics over land through the use of landfall status as an input feature. 
By coupling the input variables to both the trajectory and intensity models, a coupled storm simulation model is obtained (see Fig. \ref{ftgm}). 
The storm simulation model advances a storm's location and intensity 6 hours at a time, until one of the storm termination criteria is satisfied. 

The efficacy of the individual components of the storm simulation model, i.e., the trajectory and the intensity components, are tested by simulating five storm databases, each for a period of 100 years, which are then compared to storms from HURDAT2 since 1920. 
The synthetic storm genesis model selects a number of storms for one year based on a negative binomial distribution fitted to the number of storms each year in HURDAT2 since 1975. 
Because of enhanced storm detection, and also possibly due to global warming in this period, the average number of storms in each simulated database is 1619, compared to 1302 historical storms since 1920 in HURDDAT2.
The trajectory models faithfully represent the important statistical properties of the historical storms, such as the $6-hr$ increments in latitude and longitude (Figs. \ref{f3}, \ref{f5}, and \ref{f4a}). 
By using historical storm motion trends as input features, these models capture the dominant storm motion trends, but due to the inherent smoothing associated with LSTM-RNN models, the largest $6-hr$ increments $|\Delta \phi| > 2^\circ$, and $|\Delta \lambda| > 4^\circ$ (Figs. \ref{f3} and \ref{f4}) are under represented. 
However, historical storms that induce high $w_m$ typically have low translation speed $V$, and consequently low $|\Delta \phi|$ and $|\Delta \lambda|$. 
The values for $\Delta \phi$ and $\Delta \lambda$ plotted against the local coordinates $\phi$ or $\lambda$ in Fig. \ref{f5} show good qualitative agreement between simulated and historical trajectories. 
An anomaly arises in the frequency distributions of $\Delta \phi$ and $\Delta \lambda$ (Fig. \ref{f4}) due to the larger number of storms in the synthetic databases. 
The PDFs accounting for the larger number of synthetic storms are in better agreement (Fig. \ref{f8}). 
Additionally, the use of a reduced database (DB-3) emphasizing the most powerful storms for training the intensity models results in synthetic storms sustaining longer for storms in the medium to high $w_m$ range, and relatively low wind-speed storms dying out sooner than historical storms (see Fig. \ref{f6}). 
The global trajectories of simulated storms originating in different sub-basins are in good qualitative agreement with their historical counterparts (Fig. \ref{f7}). 

Simulated values of $p_c$ and $w_m$ are also demonstrated to be in qualitatively good agreement with the HURDAT2 storms when considering the Atlantic basin as a whole. 
Figure~\ref{f8} shows that the distributions of central pressure, $p_c(\phi)$, and pressure deficit, $\Delta p_c(w_m)$ for the simulated storms faithfully represent the historical storms. 
The near-linear relation between $p_c$ and $w_m$ is appropriately depicted by the simulated storms. 
Spreads in the distributions of $p_c$ and $w_m$ are correctly captured by the intensity models. 
The PDFs of $w_m$ in Fig. \ref{f9} for all five simulated databases are in excellent agreement with the historical storms. 

Storms achieving $w_m \ge 40$ $knots$ have been analyzed further for three chosen cities, Miami, New Orleans, and Cape Hatteras, which are all greatly affected by storm surges. 
The cumulative exceedance frequency of $w_m$ for these cities (Figs. \ref{f10}, \ref{f12}, and \ref{f14}) show excellent agreement between the five simulated databases and the historical storms for high wind speeds $w_m > 70$ $knots$. 
The cumulative frequencies are over-predicted for $w_m$ below this limit, which is 
expected on account of the higher number of storms in the simulated databases. 
The storm trajectories of the storms passing within $100$ $km$ of these cities are shown for one of the simulated databases and compared with the historical storms in Figs. \ref{f11}, \ref{f13}, and \ref{f15}. 
The number of storms affecting these areas is larger for the simulated storms over a $100-yr$ period due to larger number of storms in the simulated databases.  
The trajectory trends of the simulated storms are also in good agreement with the historical storms; although, the trajectories of the simulated storms are noticeably smoother. 
The most violent storm from each of the six databases (five simulated and HURDAT2) within 100 $km$ of these areas is extracted, and the intensity properties, $w_m$, $V$, and $\Delta p$ are plotted in Figs. \ref{f11a}, \ref{f13a}, and \ref{f15a}. 
These figures show that the six-hourly rate changes in the simulated storms' intensity and trajectory are similar to those of the actual storm, but still evolve distinctly, highlighting the stochastic nature of the models.  The trajectory of the most intense simulated storm in each of the five simulated databases, colored by $w_m$, shows good qualitative agreement to the most intense storm in HURDAT2 when considering the spatial distribution of the high wind speeds. 

Our results also show good agreement between simulated and historical spatial distributions of the speeds $w_m$ within a storm, and that the synthetic storm simulation model is capable of replicating the statistical properties of the HURDAT2 storms globally (Sec. \ref{sec3}), locally (Sec. \ref{sec4}), and over relatively short periods of time $\approx 100$ $yrs$.  The simulation model is able to capture the global statistical properties by maintaining adequately the local storm characteristics through appropriate representation of the $6-hr$ evolution of an individual storm that emulates realistic storm evolution, i.e., via excellent fine-grain representation of storm evolution. 
For these reasons, this work provides the basis for using DL and ML techniques to simulate hurricanes that can be used to accurately characterize the extreme wind climate of the Gulf and East coastlines of the US.

\clearpage
\bibliographystyle{ametsocV6}
\bibliography{reference}

%

%

\end{document}